\documentclass[11pt,graphicx,amsmath]{article}
\usepackage{ulem}
\usepackage{longtable}
\usepackage{amsmath}
\usepackage{txfonts}
\usepackage{amssymb}
\usepackage{stmaryrd}
\usepackage{mathrsfs}
\usepackage{graphicx}
\usepackage{subfigure}
\usepackage{bm}
\usepackage{float}
\usepackage{slashbox}
\usepackage[numbers,sort&compress]{natbib}
\usepackage[dvips]{color}

\usepackage{color}

\def\gsim{\;\rlap{\lower 2.5pt  \hbox{-}}\raise 1.5pt\hbox{$>$}\;}
\def\lsim{\;\rlap{\lower 2.5pt  \hbox{-}}\raise 1.5pt\hbox{$<$}\;}
\def\edth{\;\raise1.0pt\hbox{$'$}\hskip-6pt\partial\;}
\def\baredth{\;\overline{\raise1.0pt\hbox{$'$}\hskip-6pt \partial}\;}

\def\bl#1\el{\begin{align}#1\end{align}}
\def\be{\begin{equation}}
\def\ee{\end{equation}}
\def\ba{\begin{eqnarray}}
\def\ea{\end{eqnarray}}
\def\nn{\nonumber}

\def\l{\left}
\def\r{\right}

\title{  Estimation of spectrum and parameters of relic gravitational waves
             using space-borne interferometers  }

\author{\small Bo Wang \thanks{ymwangbo@mail.ustc.edu.cn },
        Yang Zhang  \thanks{yzh@ustc.edu.cn}    \\
     \small {\it Key Laboratory for Researches in Galaxies and Cosmology}, \\
     \small {\it Department of  Astronomy,  University of Science and Technology of China}, \\
     \small {\it Hefei, Anhui, 230026,  China}\\     }
\date{}

\begin{document}
\maketitle
\begin{abstract}
{\large

We present a  study of   spectrum estimation of relic gravitational waves (RGWs)
as a Gaussian stochastic background
from  output signals of future space-borne interferometers,
like  LISA and ASTROD.
As the  target of detection,
the analytical spectrum of RGWs generated during inflation
is described by three parameters:
the tensor-scalar ratio, the spectral index and the running index.
The Michelson interferometer is shown to have a better sensitivity
than Sagnac, and symmetrized Sagnac.
For RGW detection,
we analyze the auto-correlated signals  for a single interferometer,
and the cross-correlated, integrated as well as un-integrated signals
for a pair of interferometers,
and give the signal-to-noise ratio (SNR) for RGW,
and obtain lower limits of the RGW parameters that can be detected.
By suppressing noise level,
a pair has a sensitivity  2 orders better than a single
for one year observation.
SNR of LISA will be 4-5 orders higher than
that of Advanced LIGO for the default RGW.
To estimate the spectrum,
we adopt  the maximum likelihood (ML) estimation,
calculate the mean and covariance of signals,
obtain the Gaussian probability density function (PDF)
 and the likelihood function,
and derive expressions for the Fisher matrix
and the equation of the ML estimate for the spectrum.
The Newton-Raphson method is used to solve the equation by iteration.
When the noise is dominantly large,
a single LISA is not effective for estimating the  RGW spectrum
as the actual noise in signals is not known accurately.
For cross-correlating a pair, the spectrum can not be estimated from
the integrated output signals either,
and only one parameter
can be estimated with the other two being either fixed or marginalized.
We use the ensemble averaging method to estimate the RGW spectrum
from  the un-integrated output signals.
We also adopt a correlation of un-integrated signals
to estimate the spectrum and three parameters of RGW in a Bayesian approach.
For all three methods,
we provide  simulations to illustrate their feasibility.

}
\end{abstract}


\section{Introduction}

\Large

Gravitational waves (GWs) are a prediction of Einstein's
theory of general relativity,
and have been the subject of  theoretical study and continuous  detection hunting.
There are two kinds of GW, i.e,
the first  includes those  generated by  astrophysical  processes
such as inspiral of compact  binaries,
merging of massive black holes,  super-massive black hole binaries \cite{JanssenHobbs2014}, etc.
The frequencies of these sources are
typically in the range $f\sim 10^{-9} - 10^3 $\,Hz.
Examples are GW150914, GW151226 and  GW170104 from merging of binary black holes
and  GW170817 from  the binary neutron star inspiral
that was recently reported by
Advanced LIGO and Advanced Virgo as the first direct detections  \cite{GW150914,PRL2017LIGO,GW170817}.

Another kind is  the relic gravitational wave (RGW),
which is generated  during the inflation stage of cosmic expansion,
as generically  predicted by inflation models
\cite{Grishchuk1975}.
 RGW carries crucial information about the very early Universe,
such as the energy scale and slope of inflation potential,
the initial quantum states  during inflation
   \cite{zhangyang05,WangZhangChen2016},
as well  as the reheating process  \cite{TongZhang2014}.
This is because,
to the linear level of metric perturbations,
 RGW is independent of other matter components
 and its propagation is almost free.
The influences due to neutrinos free-streaming \cite{Weinberg2004,MiaoZhang2007},
quark-hadron transition and $e^+e^-$ annihilation
are minor  modifications \cite{WangZhang2008}.
This is in contrast to the scalar metric perturbation,
which is always coupled to cosmic matters
and whose short wavelength modes have gone into nonlinear evolution at present.
The second-order perturbation beyond the linear perturbation
has been also studied for RGW
\cite{AnandaClarksonWands2007,WangZhang2017}.

RGW has several    interesting  properties that are quite valuable for GW detection.
It is a stochastic background of spacetime fluctuations
distributed everywhere  in the present Universe, just like
the cosmic microwave background (CMB).
Moreover,  RGW  also exists  all the time,
and its spectrum changes very slowly on a cosmic time-scale,
so that its detections can be repeated at any time,
in contrast to  short-duration GW radiations events such as  merging of binaries.
Another  big plus of RGW for detections is that
its frequency range is extremely  broad,
stretching over  $f\sim  10^{-18}  - 10^{11}$\,Hz.
Thus, RGW is a major target for various kinds of GW detectors
at various  frequency bands,
using various  technologies \cite{TongZhang2009},
such as
CMB anisotropies and polarization measurements ($ 10^{-18}-10^{-16} $\,Hz),
   by COBE \cite{COBESmoot1992},
   WMAP \cite{WMAP1,Hinshaw2003Apjs},
   Planck \cite{Plancl2016Overview,Plancl2016Spectrum},  etc,
pulsar timing arrays ($  10^{-9}-10^{-7}$\,Hz)  \cite{PTA},
space laser interferometer ($ 10^{-5}-10^{0} $\,Hz),
  for  LISA  \cite{Bender1998,Danzmann1997,Estabrook2000},
  ($ 10^{-6}-10^{0} $\,Hz) for ASTROD   \cite{ASTROD}
and  for Tianqin and Taiji \cite{TianqinTaiji},
ground-based  laser interferometers   $(10-2000$\,Hz),
like LIGO \cite{LIGO,LIGO and VirgoS6,LIGOnoise},
    Virgo \cite{VIRGO}, GEO \cite{GEO}, KAGRA \cite{KAGRA}   etc,
cavity detectors ($\sim 4000$\,Hz) \cite{MAGO2005},
waveguide detectors  ($\sim 10^{8}$\,Hz) \cite{WaveguideCruise} and
polarized laser beam detectors ($\sim  10^{10}$\,Hz)
          \cite{Li2003}.

A primary  feature of the RGW spectrum is that
it has higher amplitude at lower frequencies \cite{zhangyang05}.
The highest amplitude is located around ($ 10^{-18}-10^{-16} $)Hz
which is the target of  CMB measurements.
So far,   the magnetic polarization $C^{BB}_l$ induced  by RGW
\cite{BaskoPolnarev1984}
has not yet been detected,
and only some  constraint is given
in terms of  the tensor-scalar ratio of metric perturbations
$r < 0.1 $  \cite{Plancl2016Overview,Plancl2016Spectrum}.
On the other hand, Advanced LIGO-Virgo
so far has not detected  RGW,
but rather has only been applied to
predict a total stochastic GW background with amplitude
$1.8^{+2.7}_{-1.3}\times10^{-9}$ near 25\,Hz
contributed together by unresolved binaries, RGW, etc \cite{Abbottetal2018}.
In between is the   band  of the space-borne facilities,
LISA and ASTROD,
where  the  amplitude of
RGW  is higher by 5-6  orders than that in the LIGO band.
This great enhancement increases
the chance for space-borne interferometers  to detect RGWs
if their sensitivity level is comparable to LIGO.

In this paper,
we  shall study RGW  detection
by  space-borne interferometers, such as LISA and ASTROD and the like,
and show how to estimate the spectrum and parameters of RGW
from  output signals of  future observations.
For this purpose,
we shall first briefly introduce the theoretical RGW  spectrum
as a scientific target,
resulting from an analytical solution that covers
from inflation to the present acceleration \cite{zhangyang05}.
Accurate estimation of  this spectrum will also
confirm the details of inflation for the very early Universe.
In this sense, this will be a direct detection of inflation.
For the RGW spectrum in this paper,
we focus on three parameters determined by inflation:
the tensor-scalar ratio $r$, the spectral index $\beta$
and the spectral running index $\alpha_t$
 \cite{TongZhang2009,WangZhangChen2016}.
Small modifications of RGW
in Refs.\cite{Weinberg2004,MiaoZhang2007,WangZhang2008}
are not considered.
We do not consider the Doppler modulation due to orbital motion,
or related causes \cite{LISACornishLarson2003}.
One of the main obstacles to detecting RGW is
the stochastic foreground of a GW resulting
from the superposition of
a large number of unresolved astrophysical sources.
To have a definite model of the power spectrum of
the stochastic foreground,
one has to know the spectrum for each type of source,
as well as  the evolution of each type.
There are several categories of these source types.
Due to the large  abundance,
Galactic white dwarf binaries are generally considered
one of the main components of the foreground  in the $\sim10^{-3}$\,Hz frequency band \cite{WhiteDwarfsForeground}.
Refs.~\cite{ForegroundModels,SetoCooray200470,AdamsCornishLittenberg2012,AdamsCornish2014}
provide several models of the foreground
generated by distribution of these binaries.
Ref.~\cite{SesanaHaardt2004} have studied
a stochastic foreground
from massive black hole binaries and its contribution to the LISA datastream.
Ref.~\cite{AMCNvBinary} show the possibility of a foreground
generated by an AM CVn binary system.
To explore the effects of these foreground models
on a spaced-based detector,
simulation methods to generate a foreground data stream for LISA
have been studied by the Mock LISA data challenge
project \cite{MockDataChallenge} and other groups \cite{CornishCrowder2005,CornishRobson2017}.
Based on these dummy datastreams,
several techniques for model selection and parameter estimation
have been developed \cite{SetoCooray200470,RobsonCornish2017,CornishLittenberg2007,CornishRobson2017,AdamsCornish2014}.
Refs.~\cite{CrowderCornish2007prd,RobsonCornish2017} have provided
methods to detect resolvable sources in a foreground
of unresolvable sources.
Ref.~\cite{AdamsCornish2014} investigated
approaches to discriminate the GW background from a stochastic foreground
according to the differences in spectral shapes
and time modulation of the signal.
Currently,
the foreground is still under intensive study but is not fully understood.
At this stage of our study
we do not include the foreground in this paper.

GW radiation from a finite   source
usually has  a definite waveform (fixed direction, amplitude, etc)
and  the match-filter method \cite{Helstrom1960}
is  commonly used to estimate the waveform
against certain theoretical templates.
Refs.\cite{Cutler & Flanagan1994,Finn1992} studied the methods of
parameter estimation  for ground-based  LIGO detectors.
Refs.\cite{Cutler1998,MooreHellings2002} studied
   detection of a GW radiated from merging compact binaries using LISA detectors.
In contrast, RGW is of stochastic nature, incident from all directions,
containing modes of all possible frequencies and amplitudes.
Refs.\cite{Flanagan1993, Allen & Romano}
systematically studied detection of RGW using  LIGO,
obtained a formula for signal-to-noise ratio (SNR)
as  a criterion for detection.
Ref.\cite{Binetruy2012}  discussed
the possible detection by eLISA
of GW backgrounds due to first-order phase transitions, cosmic strings,
   bubble collision, etc.
Ref.\cite{UngarelliVecchio2001} discussed possibility of RGW detection by LISA.
So far  in literature, however, RGW detection by space-based interferometers
has not been systematically studied,
in particular,  estimation for the RGW spectrum   has not been analyzed.
We shall  derive  formulations of estimation of the RGW spectrum,
using a single or a pair of space-based interferometers like LISA and ASTROD etc.

For this purpose,
we shall briefly  examine  the three kinds of interferometers:
Michelson,  Sagnac and symmetrized Sagnac
\cite{Vine,Cornish & Rubbo,LISAShaddock2003,
Estabrook2000,Cornish & Hellings, Schilling,Vallisneri,
Larson & Hiscock, Cornish,Cornish & Larson,RobinsonRomano2008},
whose sensitivity   depends on both the noise
and transfer function,
which in turn depends on the detector geometry.
We shall show explicitly that
the  Michelson  has the best sensitivity,
which will be taken as a default interferometer.
For  a single interferometer in space,
we give SNR and a criterion to detect RGW.
As a Gaussian stochastic background,
RGW is similar to CMB  anisotropies, and the statistical methods
employed in CMB studies can be used
\cite{Gorski1994,Oh1999,Hinshaw2003Apjs}.
We shall apply the maximum likelihood (ML) method    \cite{Kay}
to  estimate the RGW spectrum.
We give the probability density function (PDF) explicitly,
and derive the estimation equation of an RGW spectrum.
However, in practice, our knowledge of the spectrum
of noise that is actually occurring in the detector
is not sufficient
so that  a single case is not effective to estimate the RGW spectrum
when the noise is dominantly large.

For a pair,
the noise level will be suppressed by  cross-correlation.
We shall introduce cross-correlated, integrated output of the pair,
in a  fashion similar to the ground-based LIGO  \cite{  Allen & Romano},
calculate the overlapping reduction function,
give the sensitivity and compare with that of a single case,
and analyze possible detection and constraints on RGW parameters.
However, the spectrum as a function of frequency can not be estimated
from the   integrated output,
since the   frequency-dependence has been lost in integration.
One can estimate only one parameter in the  Bayesian approach
by  ML-estimation,
  using the Newton-Raphson method \cite{Oh1999,Hinshaw2003Apjs,Press1992}.
To estimate the spectrum,
we propose the ensemble averaging method,
and directly take the cross-product of un-integrated output signals from a pair.
The method does not depend on precise knowledge of the noise spectrum.
We estimate the spectrum using simulated data for illustration,
but one can not estimate the parameters.
Ref. \cite{Seto2006} suggested a method of correlation for un-integrated signals,
by which the whole frequency range of the data is to be divided
into many small segments  of frequency,
and the mean value of a correlation variable over each small segment
is taken as the representative point for the segment.
In this way, as an approximation,
the correlation variable  as a function of frequency
is defined on the whole range.
Ref. \cite{Seto2006} considered a simple power-law spectrum of stochastic GW,
analyzing the resolution of parameter estimation,
but did not give an estimation of the spectrum.
We adopt this as the  third method
to estimate the RGW spectrum by ML-estimation,
as well as the three parameters  ($r$, $\beta$, $\alpha_t$)  simultaneously
 in a  Bayesian approach.
For all these three methods for a pair,
we shall  provide numerical simulations,
demonstrating their  feasibility.

The outline of the paper is as follows.
In Section 2, we give a short review of
the theoretical RGW spectrum.
In Section 3 we compare briefly  the sensitivity of
three types of interferometers,
and give a constraint on RGW by a single Michelson in space.
In Section 4,
we examine signals from a single by the ML method
and show that it is not effective to estimate the RGW spectrum
when the noise is dominantly large.
Section 5 is about the cross-correlated, integrated output signals for a pair.
In Section 6,
 we show that the integrated output signals from a pair
can be used to estimate one parameter, but not the spectrum.
In Section 7 we use the ensemble average  method
to  estimate the spectrum directly.
In Section 8, we use  the correlation method for un-integrated output signals
to  estimate the spectrum and parameters of RGW.
Appendix A  gives the  derivation of the Fisher matrix   for  a pair.

\section{  Relic Gravitational Wave}

\begin{figure}[htbp]
\centering
 \includegraphics[width=0.6\linewidth]{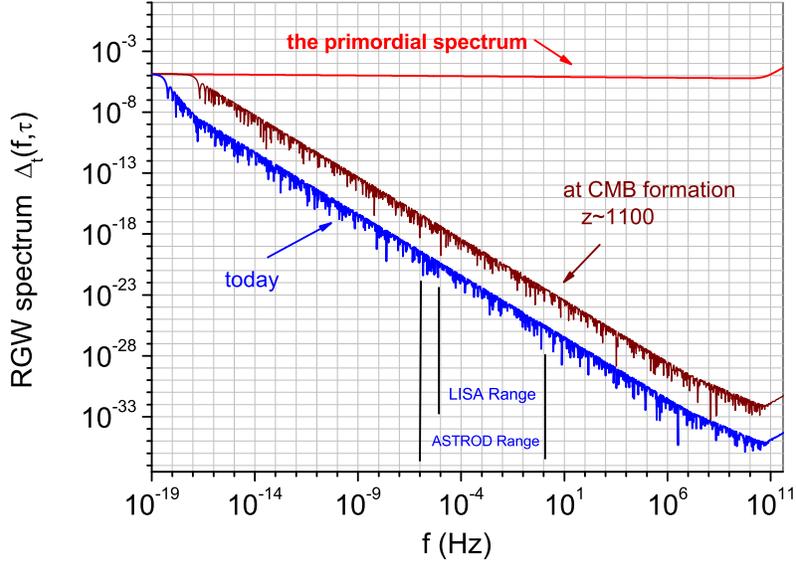}
\caption{The evolution of RGW spectrum
   from  inflation to the present.
   }\label{evolspectr}
\end{figure}

This section reviews the main properties of RGWs relevant to detection by LISA.
RGW   as the tensor metric perturbations of spacetime is generated
during inflation and
exists as a stochastic background of fluctuations in the Universe.
It  has an  extremely broad spectrum,
ranging from $10^{-18}$\,Hz to $10^{11}$\,Hz.
In particular, it has a characteristic amplitude of $10^{-22} \sim 10^{-24}$
around $f\sim   10^{-3}$\,Hz (see Fig.\ref{evolspectr})
and can serve as a  target for LISA.
The exact solution
and corresponding analytical spectrum of RGW
have been obtained  \cite{zhangyang05,WangZhangChen2016}
 that cover the whole course of expansion,
from inflation, reheating, radiation, matter, to the present accelerating stage.

For a spatially flat   Robertson-Walker spacetime,
 the metric with tensor perturbation  is
\be \label{RW metric}
ds^2=a(\tau)^2[-d\tau^2+(\delta_{ij}+h_{ij})dx^idx^j]
\ee
where $h_{ij}$ is the tensor perturbation
and $\tau$ is the conformal time.
From the inflation to the accelerating expansion,
there are five stages, with each stage being described by
a power-law scale factor  $a(\tau) \propto \tau^d$
where $d$ is a constant \cite{zhangyang05}.
The particularly interesting stage is inflation with
\begin{equation}\label{inflation stage}
a(\tau)=l_0|\tau|^{1+\beta},
\hspace{10mm}
-\infty<\tau\leq\tau_1,
\end{equation}
where  $\beta  $ is the spectral index.
For the exact de Sitter,  $\beta=-2$, and for generic inflation models
and $\beta $  can deviate slightly  from -2
 \cite{zhangyang05}.
The present accelerating stage has
\be \label{acc Stage}
a(\tau)=l_H|\tau-\tau_a|^{-\gamma },
\hspace{10mm}
\tau_E\leq\tau\leq\tau_H,
\ee
where   $\gamma =2.018$ is taken for $\Omega_\Lambda=0.71$ \cite{WangZhangChen2016}.
The normalization is taken as
  $a(\tau_H)=  l_H = \gamma /H_0$,
where $H_0$ is the present Hubble constant.

The tensorial perturbation $h_{ij}$ as a  quantum  field
is decomposed into the Fourier modes,
\begin{equation}\label{h Fourier}
h_{ij}  ({\bf x},\tau)=\int\frac{d^3k}{(2\pi)^{3/2}}
    \sum_{A ={+,\times}}  \epsilon^A_{ij}({\bf k})
    \left[ a^A_{\bf k}   h^A_k(\tau) e^{i\bf{k}\cdot\bf{x}}
    +a^{A\, \dagger}_{\bf k}h^{A\, *}_k(\tau)e^{-i\bf{k}\cdot\bf{x}}\right]
       , \,\,\,\, {\bf k}=k\hat{k},
\end{equation}
where $a^A_{\bf k}$ and $a^{A\, \dagger}_{\bf k}$ are
the annihilation and creation operators respectively
of a graviton with
wavevector $\bf k$ and polarization $A$,
satisfying  the   canonical commutation relation
\be \label{aadagger}
 \left[a^A_{\bf k},\,   a^{A' \, \dagger}_{\bf k'}\right]
    =  \delta_{A A'} \,  \delta^3({\bf k-k'}).
\ee
Two polarization tensors satisfy
\be\label{polariz}
 \epsilon^A_{ij}({\bf k}) \delta_{ij}=0,\,\,\,\,
 \epsilon^A_{ij}({\bf k})  k^i=0, \,\,\,\,
 \epsilon^A_{ij}({\bf k})  \epsilon^{A'}_{ij}({\bf k}) =2\delta_{AA'},
\ee
and can  be taken as
\[
\epsilon_{ij}^+ ({\bf k})= (l_i l_j - m_i m_j),
\,\,\,\,
\epsilon_{ij}^\times ({\bf k})= ( l_i m_j + m_i l_j) ,
\]
where $\bf l$, $\bf m$ are mutually  orthogonal unit vectors normal to $\bf k$.
In fact, as an observed quantity for LISA,
RGW can be also treated as a classical,  stochastic field
\be \label{hFourierclass}
h_{ij}(\tau,x)=\int \frac{d^3k }{(2\pi)^{3/2}}  \sum_{A
          =\times,+} \epsilon^A_{ij}({\bf k}) h^{A}_k(\tau) e^{i \mathbf{k\cdot x}},
\ee
where the $k$-mode $ h^{A}_k $ is  stochastic,
independent of other modes.
The physical frequency  at present is related to
  the conformal wavenumber via
$f= ck/2\pi a(\tau_H) $   \cite{zhangyang05}.
For RGW,
the two  polarization modes $h^+_k$ and $h^{\times}_k$
are assumed to be independent and  statistically  equivalent,
so that the superscript $+, \times$ can be dropped,
and the wave equation   is
\be  \label{GWs equation}
h''_k(\tau)+2\frac{a'(\tau)}{a(\tau)}h'_k(\tau)+k^2h_k(\tau)=0 .
\ee
The   quantum state during inflation
 is taken  to be $|0 \rangle$ such that
\be\label{ask}
 a^s_{\bf k}  |0 \rangle =0, \,\,
\ee
i.e,
only the vacuum fluctuations   of RGW
are  present during inflation,
and the solution of RGW is
\be  \label{sol}
h_k(\tau)
=\frac{\sqrt{32\pi G}}{ a(\tau)}   \sqrt{ \frac{\pi k|\tau| }{2k}}
              \left(-ie^{-i\pi \beta/2}\right) H^{(2)}_{\beta+ \frac{1}{2}} (k|\tau|)
,\,\,\, \, -\infty<\tau\leq \tau_1 \,
\ee
which  is the positive-frequency mode
$h_k \rightarrow  \frac{\sqrt{32\pi G}}{a(\tau)}
\frac{1}{\sqrt{k}}e^{-ik\tau}$ and
gives a zero point energy $\frac{1}{2}\hbar \omega$
in each $\bf k$-mode and each polarization
in the high frequency limit.
The wave  equation  (\ref{GWs equation}) has been solved also
for other subsequent stages, i.e,
reheating, radiation dominant, matter dominant and accelerating.
The solution of Eq.~(\ref{GWs equation})
is simply a combination of two Hankel  functions,
$\tau^{d-1/2} H^{(1)}_{d-1/2}$ and $\tau^{d-1/2}H^{(2)}_{-d+1/2}$.
By continuously joining these stages,
the full analytical solution $h_k(\tau)$ has been obtained,
which covers the whole course of evolution,
in particular, for the present stage of acceleration,
it  is given by   \cite{WangZhangChen2016}
\be \label{upresent}
h_k(\tau ) = \frac{\sqrt{32\pi G}}{a(\tau)}  \sqrt{\frac{\pi s}{2k}}
     \bigg[e^{-i\pi\gamma/2}\beta_k  H^{(1)}_{-\gamma-\frac{1}{2} } (s)
          +e^{i\pi\gamma/2}\alpha_k  H^{(2)}_{-\gamma-\frac{1}{2}} (s) \bigg],
         \,\,\,\,\, \, \tau_E <\tau\leq \tau_H,
\ee
where $s \equiv k(\tau-\tau_a)$ and
the coefficients $\beta_k, \alpha_k$ are Bogoliubov coefficients
\cite{ParkerToms}
satisfying $|\alpha_k |^2- |\beta_k |^2 =1$,
and $|\beta_k|^2$ is the number of gravitons at the present stage,
and the    expressions $\beta_k, \alpha_k$
are explicitly given  by Ref.\cite{WangZhangChen2016}.
The frequency range of space-borne interferometers
is much higher than
the Hubble frequency $ H_0 \simeq  2\times 10^{-18}$ Hz,
so that (\ref {upresent}) for these modes  becomes
\be
h_k(\tau ) \simeq \frac{\sqrt{32\pi G}}{a(\tau)}
\frac{1}{\sqrt{k}}e^{-ik\tau} ~~ for ~~k ~~ \gg 1/ |\tau|  .
\ee
Hence, for space-borne interferometers,
RGW is  practically
a superposition of stochastic plane waves.

The auto-correlation function of RGW
is defined by the following expected value
\ba\label{vevcorr}
\langle0|  h^{ij}(\textbf{x},\tau)h_{ij}(\textbf{x},\tau) |0\rangle
         =   \frac{1}{(2\pi)^3} \int d^3k \,  | h_k |^2    ,
\ea
where   (\ref{polariz})  (\ref{aadagger})  have been used.
Defining  the   power spectrum by
\be \label{defspectrum}
\langle0|  h^{ij}(\textbf{x},\tau)h_{ij}(\textbf{x},\tau) |0\rangle
  \equiv
\int_0^{\infty}\Delta^2_t(k,\tau)\frac{dk}{k} \, ,
\ee
one reads off the   power spectrum
\be \label{spectrum}
\Delta^2_t(k,\tau)= \frac{ k^{3}}{2\pi^2}|h_k(\tau)|^2 ,
\ee
which is dimensionless.
We also use a notation $h(f,\tau_H )\equiv \sqrt {\Delta^2_t(k,\tau_H) }$.
In the literature  on GW detection,
the characteristic amplitude
\be\label{charampl}
h_c(f) \equiv  \frac{ h(k,\tau_H)}{  2\sqrt{f}}
\ee
is often used \cite{Maggiore,Zhang2010}, which has dimension Hz$^{-1/2}$.
The  definition (\ref{spectrum}) holds
for any time $\tau$,  from inflation to the  accelerating stage.
Fig.~\ref{evolspectr} shows the evolution of the RGW spectrum from inflation
to the present acceleration stage.
Equivalently, one can also use the spectral energy density
  $\Omega_g \equiv \rho_g/\rho_c$,
where
\[
\rho_g=\frac1{32\pi G a^2}
\langle 0|  h'_{ij }\,  h'^{ij} |0 \rangle
\]
is the energy density of RGW \cite{SuZhang,WangZhangChen2016}
and $\rho_c=3H_0^2/8\pi G$ is the critical density.
The spectral energy density $\Omega_g(f)$
is defined by $\Omega_g \equiv \int\Omega_g(f)df/f$,
 and given by
\be\label{energy density of RGW}
\Omega_g(f)=\frac{\pi^2}3 h^2(f,\tau_H) \l(\frac f{H_0}\r)^2,
\ee
which holds for all wavelengths    shorter than the horizon.

From the  spectrum (\ref{spectrum})  during inflation,
the analytic expressions of spectral, and running spectral indices
have been obtained  \cite{WangZhangChen2016}
$n_t  \equiv\frac{d\ln\Delta_t^2}{d\ln k}
\simeq  2\beta+4 -\frac{2}{2\beta+3 }x^2 $,
$\alpha_t  \equiv \frac{d^2\ln \Delta^2_t}{d(\ln k)^2}
\simeq  -\frac{4}{2\beta+3 } x^2$
at $x \equiv |k\tau| \ll 1$, i.e, at far outside horizon,
both related to the spectral  index $\beta$.
In the limit $k \rightarrow 0$,
one has the default values
\be
n_t=  2\beta+4 \, , \,\,\,\, \alpha_t = 0,
\ee
which hold for the inflation models
 with $a(\tau) \propto |\tau|^{1+\beta}$.
It is incorrect  to use  $n_t$ and $\alpha_t$
evaluated at the horizon-crossing $|k\tau| =1$
   \cite{KosowskyTurner1995}.
With these definitions,
the primordial spectrum in the limit $k \rightarrow 0$ is written as
\be \label{initialSprctrum}
\Delta_t (k)
=\Delta_{R} \, r^{1/2}(\frac{k}{k_{0}})^{\frac{1}{2}n_t
+\frac{1}{4}\alpha_t \ln(\frac{k}{k_0})},
\ee
where $k_{0}$ is a pivot conformal wavenumber
  corresponding to a physical wavenumber
$k_0/a(\tau_H) = 0.002 $Mpc$^{-1}$, $\Delta_{R}$ is the value of
curvature perturbation determined by observations
  $\Delta_{R}^{2}=(2.464\pm 0.072)\times10^{-9}$,
and $r\equiv \Delta^2_{t}(k_0)/\Delta^2_{R}(k_0)$ is the tensor-scalar ratio,
and  with $r< 0.1$ by CMB observations \cite{Plancl2016Spectrum}.
The primordial spectrum (\ref{initialSprctrum}) describes
the upper curve (red)  during inflation   in  Fig.~\ref{evolspectr}.
The present  spectrum $\Delta_t (f,\tau_H )$  and
the primordial spectrum $\Delta_t(k)$
are overlapped at  very low frequencies $f < 10^{-18}$ Hz,
with both being $ \propto r^{1/2}$ as in (\ref {initialSprctrum}).
At $f> 10^{11}$\,Hz,
$\Delta_t (f,\tau_H )$ rises up and has a UV divergence,
due to  vacuum fluctuations.
In Ref.\cite{WangZhangChen2016},
the UV divergence has been adiabatically  regularized.
Higher values of ($r$, $n_t$, $\alpha_t$) give rise to
higher amplitude of RGW.
In particular, a slight increase in  $\alpha_t$ will
enhance greatly the amplitude of RGW in the relevant band.
In this paper, we take  ($r$, $n_t$, $\alpha_t$) as the major parameters of RGW.

\section{Sensitivity of one interferometer and RGW Detection}

\begin{figure}[htbp]
\centering
  \includegraphics[width=0.8\columnwidth]{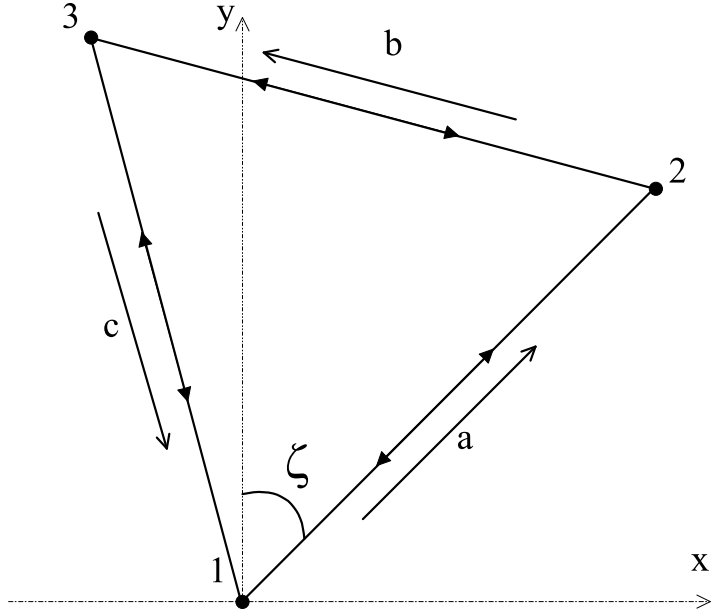}\\
  \caption{The three spacecrafts are located at points 1, 2 and 3,
and  the vectors \textbf{a},\textbf{b} and \textbf{c}
   label the direction of the three arms.}\label{figure structure}
\end{figure}

We briefly review    detection of RGW by a single interferometer,
which has been studied before and will be used in this paper later.
Figure \ref{figure structure} shows three identical spacecrafts
that are placed in space, forming an equilateral triangle.
The three arms are of equal length,
taken to be $L=5\times10^{9}$\,m by the original design of LISA.
when no GW is passing by \cite{Bender1998}.
This value is taken as an example in our paper.
In recent years,
the designed arm-length has been modified to be
$L=1\times10^{9}$\,m \cite{AmaroSeoane2012}
or  $L=2.5\times10^{9}$\,m \cite{AmaroSeoane2017}.
The recently-proposed projects,
like Tianqin and Taiji, also will have $L$ around this value.
ASTROD has a longer value of $L=260 \times10^{9}$\,m \cite{ASTROD}.
Spacecraft 1 can shoot laser beams,
which are
phase-locked, regenerated with the same phase
at the spacecrafts 2 and 3,
and then sent back \cite{AmaroSeoane2017,Bender1998}.
This forms one interferometer.
In the presence of a GW,
the arm lengths and phases of the beams   will fluctuate.
Combining the optical paths will produce different interferometers
           \cite{Vallisneri&Crowder}.
Here we only discuss three combinations,
the Michelson, the Sagnac
and the symmetrized Sagnac \cite{Cornish, Cornish & Larson}.
To focus on the main issue of spectral estimation for RGW,
we do not consider the  spacecraft orbital effects,
Shapiro delay,  etc,
caused by the Newtonian potential of the solar system
\cite{Rubbo2004,Cornish & Rubbo,Tinto1999}.

\subsection{The response tensors for three kinds of interferometers}

First, the Michelson interferometer  \cite{Cornish,Cornish & Larson,Cutler1998,Maggiore}
is considered.
The  optical path difference
is proportional to the  strain
\be \label{Michelson strain general}
h_o =\frac{1}{2L}
\left[l_{12}
+l_{21}
-l_{13}
-l_{31}\right] ,
\ee
where
 $l_{12} $ is  the optical path
of a photon emitted by spacecraft 1 traveling along arm 1-2,
which has  arrived at 2,
 $l_{21} $ is the one reflected at 2 and back to 1, etc.
This is the dimensionless strain   measured by the interferometer,
 also called the output response.

Next, the Sagnac interferometer \cite{LISAShaddock2003}
is described.
One optical path is 1-2-3-1,
and the other is along 1-3-2-1.
The strain is proportional to their difference
\be \label{sagnac signal}
h_{os1} = \frac{1}{3L}\left[l_{13}+l_{32} +l_{21}
    -l_{12}-l_{23} -l_{31}\right],
\ee
where the subscript ``1" in $h_{os1} $ refers to vertex 1.
In a similar fashion,  one can get the output responses
at   vertex 2 and  vertex  3.
Last, the symmetrized  Sagnac interferometer is examined.
Its  output response   is defined as
the average for three Sagnac signals of three vertices, 1, 2 and 3,
respectively \cite{Armstrong,RobinsonRomano2008}
\begin{equation}\label{sss}
    h_{oss}=\frac{1}{3} (h_{os1}+h_{os2}+h_{os3}).
\end{equation}
We shall  study the   ability to detect   RGW
with these three kinds of interferometers
when implemented in space.

Now we consider the responses of the three kinds of interferometers  to a GW.
Let  a plane  GW,
with frequency $f$ from a direction  $\hat\Omega$,
pass through the detector located at $\bf r$ and at  time $t$.
The GW is denoted by $\textbf{h}(\hat{\Omega},f,t,{\bf r})$ as a tensor.
The output response of an interferometer  is
a  product
\be \label{Michelson strain in-co}
h_{o}(\hat{\Omega},f,t,{\bf r})=\textbf{D}  (\hat{\Omega},f) :
    \textbf{h}(\hat{\Omega},f,t,{\bf r}) .
\ee
where $\mathbf D(\hat\Omega,f)$ is the response tensor,
depending on the orientation and geometry of LISA and the operating frequency $f$.
For the  Michelson,   the response tensor is
 \cite{Cornish & Larson, Cornish},
\begin{equation}\label{response}
\mathbf D_m(\hat{\Omega},f)=\frac{1}{2}((\textbf{a}\otimes \textbf{a})
\mathcal {T}_m (\textbf{a} \cdot \hat{\Omega},f)
-(\textbf{c}\otimes \textbf{c})\mathcal{T}_m (-\textbf{c}\cdot\hat{\Omega},f)) ,
\end{equation}
with $\bf a$ and $\bf c$ being the vectors of arms shown in Fig.~\ref{figure structure},
and  $\mathcal T_m$ being the single-arm transfer function
\bl
\mathcal{T}_m(\textbf{a}\cdot \hat{\Omega},f)
=&\frac{1}{2}
\bigg[\text{sinc}\left(\frac{f}{2f_*}(1-\textbf{a}\cdot \hat{\Omega})\right)
\exp\left(-i\frac{f}{2f_*}(3+\textbf{a}\cdot \hat{\Omega})\right)
\nn\\
&+\text{sinc}\left(\frac{f}{2f_*}(1+\textbf{a}\cdot \hat{\Omega})\right)
\exp\left(-i\frac{f}{2f_*}(1+\textbf{a}\cdot \hat{\Omega})\right)\bigg] ,\label{transfer function}
\el
where $\text{\text{sinc}} (x) \equiv\frac{\sin x}{x}$,
and $f_*\equiv c/(2\pi L)\simeq 0.0095$\,Hz  is the characteristic frequency of LISA.
In expression (\ref {Michelson strain in-co}),
 ``$:$" denotes a tensor product defined by
\[
( {\bf a}\otimes \textbf{a}):{\bm \epsilon} \equiv a_ia_j\epsilon_{ij},
\]
where  $\epsilon_{ij}$ is the  polarization tensor
satisfying the conditions of (\ref{polariz}).
The output response (\ref{Michelson strain in-co})
indicates  how the interferometer transfers an incident GW
into an output signal through a specific geometric setup.
When passing, for the ground-based case of LIGO,
the working frequency range is
 ($10^1$ $\sim$ $10^3$)Hz \cite{GW150914},
and $f_*\simeq1.2\times10^4$\,Hz for 4km-long arms,
so one can take the low frequency limit $f\ll f_*$,
$\mathcal{T}_m \simeq 1$,
then Equation (\ref{response})
reduces to that of Ref.\cite{Allen & Romano}.

 The   response tensor  of the Sagnac is \cite{Cornish,KudohTaruya2005}
\be \label{sagnac response}
    {\bf D}_s=\frac{1}{6}
    \bigg( (\mathbf{a}\otimes\mathbf{a})\mathcal{T}_a(f)
      +(\mathbf{b}\otimes\mathbf{b})\mathcal{T}_b(f)
      +(\mathbf{c}\otimes\mathbf{c})\mathcal{T}_c(f) \bigg),
\ee
depending also on the direction vector $\bf b$,
the transfer functions
\bl\label{sagnac transfer 1}
    \mathcal{T}_a(f)= & \text{sinc}\left(\frac{f}{2f_*}(1+\mathbf{a}\cdot\hat{\Omega})\right)
    \exp\left(-i\frac{f}{2f_*}(1+\mathbf{a}\cdot\hat{\Omega})\right) \nn \\
&    -\text{sinc}\left(\frac{f}{2f_*}(1-\mathbf{a}\cdot\hat{\Omega})\right)
    \exp\left(-i\frac{f}{2f_*}(5+\mathbf{a}\cdot\hat{\Omega})\right)
\el
\be \label{sagnac transfer 2}
    \mathcal{T}_b(f)=  \bigg[\text{sinc} \left(\frac{f}{2f_*}
    (1+\mathbf{b}\cdot\hat{\Omega})\right)        -\text{sinc}\left(\frac{f}{2f_*}(1-\mathbf{b}\cdot\hat{\Omega})\right)\bigg]
\exp\left(-i\frac{f}{2f_*}
    (3+\mathbf{a}\cdot\hat{\Omega}-\mathbf{c}\cdot{\Omega})\right)
\ee
\bl \label{sagnac transfer 3}
    \mathcal{T}_c(f)= & \text{sinc}\left(\frac{f}{2f_*}(1+\mathbf{c}\cdot{\Omega})\right)
    \exp\left(-i\frac{f}{2f_*}(5-\mathbf{c}\cdot{\Omega})\right)  \nn \\
&    -\text{sinc}\left(\frac{f}{2f_*}(1-\mathbf{c}\cdot{\Omega})\right)
    \exp\left(-i\frac{f}{2f_*}(1-\mathbf{c}\cdot{\Omega})\right).
\el
Notice that a factor of $\frac{1}{2}$ is missed
in  the exponent function in  (9) of  Ref.\cite{Cornish}.
The  response tensor  of the symmetrized Sagnac is  \cite{Cornish,KudohTaruya2005}
\be\label{ss response}
    \mathbf{D}_{ss}(\hat{\Omega},f)= \frac{1}{6}
    \bigg( (\mathbf{a}\otimes\mathbf{a})\ \mathcal{T}_{ss}\left(\mathbf{a}\cdot\hat{\Omega},f\right)
    +(\mathbf{b}\otimes\mathbf{b})\ \mathcal{T}_{ss}\left(\mathbf{b}\cdot\hat{\Omega},f\right)
    +(\mathbf{c}\otimes\mathbf{c})\ \mathcal{T}_{ss}\left(\mathbf{c}\cdot\hat{\Omega},f\right)
    \bigg ),
\ee
and
\bl \label{ss transfer}
    \mathcal{T}_{ss}\left(\mathbf{u}\cdot\hat{\Omega},f\right)
    =&
    \left(1+2\cos\frac{f}{f_*}\right)
    \exp\left(-i\frac{f}{2f_*}(3+\mathbf{u}\cdot\hat{\Omega})\right)
    \Bigg[\text{sinc}\left(\frac{f}{2f_*}(1+\mathbf{u}\cdot\hat{\Omega})\right)
   \nn\\
   &
    -\text{sinc}\left(\frac{f}{2f_*}(1-\mathbf{u}\cdot\hat{\Omega})\right)\Bigg].
\el
Thus, for a given incident GW,
these three kinds of interferometers
will  yield different output responses due to  response tensors.

The above output response (\ref{Michelson strain in-co}) applies to
GW emitted by a fixed source far away from the detector.
The matched filter technique is usually used
to search for a GW  embedded in the  noise
\cite{Thorne1987,Cutler & Flanagan1994,Finn1992}.

\subsection{ The output response to RGW and the transfer function}

RGW as a stochastic background contains
a mixture of all independent $\bf k$ modes of plane waves.
In regard to its  detection by space-based interferometers,
RGW  in  (\ref  {hFourierclass}) can be also written as
a sum over  frequencies and  directions \cite{Allen & Romano, Cornish}
\bl   \label{hij1}
h_{ij}(t,x)&=\sum _{A=\times, +} \int^\infty _{-\infty} df
\int d\hat{\Omega} \, \epsilon^A_{ij}(\hat{\Omega})  \ \tilde{h}_{A}(f,\hat{\Omega})
e^{-i 2\pi  f t}
e^{ i 2\pi f\hat{\Omega}\cdot\textbf{r}/c}
,
\el
where  $ \hat{\Omega}={\bf k}/k$, ${\bf{r}} =a(\tau_H)\bf{x}$,
and
\be
\tilde h_A(f,\hat\Omega)=\frac{2\pi a(\tau_H)/c }{(2\pi)^{3/2}}
    \,   k^2   h^A_k(\tau)e^{i 2\pi f t}
\ee
with the mode   $h_k^A(\tau)$ given in (\ref {hFourierclass}).
By its stochastic nature,
each mode of frequency $f$ and in direction $\hat\Omega$ is random.
Statistically,  RGW  can be assumed to be a Gaussian random process,
and  the ensemble averages  are given by   \cite{Allen & Romano}
\be \label{h emsemble}
\langle \tilde{h}_A(f,\hat{\Omega}) \rangle =0 \,  ,
\ee
\be  \label{h2 emsemble}
\langle \tilde{h}_A^*( f,\hat{\Omega})\tilde{h}_{A'}(f',\hat{\Omega '}) \rangle
   =\frac{1}{2}\delta (f-f') \frac{\delta ^2(\hat{\Omega},\hat{\Omega '})}{4\pi}
           \delta_{AA'}S_h(f) ,
\ee
where $\delta^2(\hat\Omega,\hat\Omega')
=\delta(\phi-\phi')\delta(\cos\theta-\cos\theta')$,
and $S_h(f)$ is the spectral density, also referred to as the spectrum, in the unit of Hz$^{-1}$
satisfying  $S_h(f)=S_h(-f)$.
The factor $\frac{1}{2}$ is introduced
considering
that the variable $f$ of $S_h$ ranges between $-\infty$ and $+\infty$.
Eqs. (\ref {h emsemble}) and (\ref {h2 emsemble})   specify
 fully the statistical properties of RGW.
The normalization of $S_h(f)$ is chosen such that
\bl\label{Sh normalization}
\langle \tilde{h}_A^*( f)\tilde{h}_{A'}(f') \rangle
&= \int d\hat \Omega d\hat\Omega'\langle
\tilde{h}_A^*( f,\hat{\Omega})\tilde{h}_{A'}(f',\hat{\Omega '}) \rangle\nn
\\
&=\frac12\delta(f-f')\delta_{AA'}S_h(f).
\el
From Eqs. (\ref{hij1}) and (\ref{h2 emsemble}),
the auto-correlation function of RGW can be written as
\be\label{spectral density2}
\langle h_{ij}(t)h^{ij}(t)\rangle=2\int_{-\infty}^{+\infty}df S_h(f)
   =4   \int_0^\infty d(\log f)f S_h(f),
\ee
so that  the spectral density  is related to the characteristic amplitude
(\ref   {charampl} )  as the following
\be \label{hspectrum}
  S_h(f) = h^2_c(f)
         =\frac{3H_0^2}{4\pi^2}\frac{\Omega_g(f)}{f^3} \, ,
\ee
where $\Omega_g(f)$  is
the spectral energy density (\ref{energy density of RGW}).

\begin{figure}[htbp]
\centering
  \includegraphics[width=0.5\columnwidth]{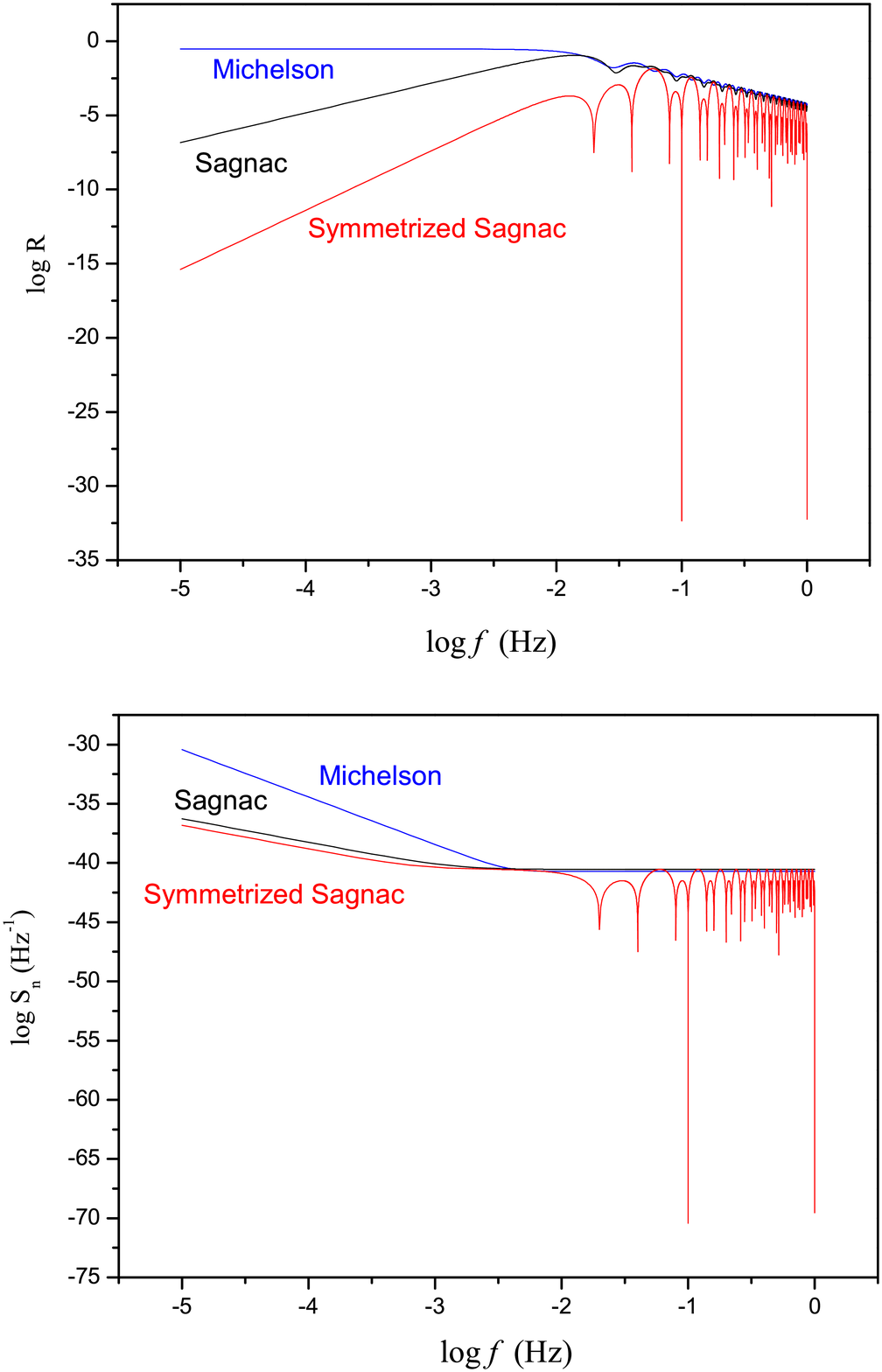}
  \caption{
  Top: the transfer functions $\mathcal{R}(f)$.
  Bottom: the noise spectrum $S_n(f)$.
  }\label{R+S}
\end{figure}

Let us consider the output response of an interferometer to  RGW.
Substituting $h_{ij}$ of Eq.~(\ref{hij1}) into
Eq.~(\ref{Michelson strain in-co})
yields the output response
\be \label{sm expansion}
h_o(t)= \sum_{A=\times, +}\int_{-\infty}^{+\infty}df\int d\hat \Omega \,
   \tilde h_A(f,\hat\Omega)e^{-i2\pi ft}
   e^{i2\pi f\hat{\Omega}\cdot\textbf{r}/c} \,
  \mathbf D(\hat\Omega,f) :\mathbf \epsilon ^A(\hat\Omega) ,
\ee
which is valid for all three kinds
of interferometers with
their respective response tensor $\mathbf D $.
Since the RGW background is isotropic in the Universe,
we are free to take the detector location at $\bf r =0$  \cite{Cornish & Larson},
so that Eq.~(\ref{sm expansion}) becomes
 \be \label{strain}
 h_o(t)=\sum_{A=\times, +}\int_{-\infty}^{+\infty}df\int d\hat
 \Omega\tilde h_A(f,\hat\Omega)e^{-i2\pi ft}  \,
 \mathbf D(\hat\Omega,f):\mathbf \epsilon ^A(\hat\Omega) ,
 \ee
which is a summation over all frequencies, directions and polarizations,
in contrast to a GW from a fixed source.
Its  Fourier transform   is
\be \label{hij}
\tilde h_o(f)=\sum _A
\int d\hat{\Omega}\ \tilde{h}_{A}(f,\hat{\Omega})
\textbf{D}(\hat{\Omega},f): \epsilon^A(\hat{\Omega}) .
\ee
 The  ensemble averages  (\ref{h emsemble})  (\ref{h2 emsemble}) lead to
\bl\label{Sh normalization2}
\langle \tilde{h}_o(f) \rangle=0,
    ~~~~
\langle \tilde h^*_o(f)\tilde h_o(f') \rangle
=\frac12\delta(f-f')S_h(f)\mathcal{R}(f).
\el
The auto-correlation of output response is
\begin{equation}\label{sm ensemble average}
\langle h_o^2(t) \rangle
=\int^{+\infty}_{-\infty} df\  \frac12 S_h(f) \mathcal{R}(f)
= \int ^\infty _0 df\ S_h(f) \mathcal{R}(f) ,
\end{equation}
where   the transfer function
\be \label{transferGeneral}
\mathcal{R}(f) =\int\frac{d\hat{\Omega}}{4\pi}
    \sum_{A=\times, +}F^{A*}(\hat{\Omega},f)\ F^{A}(\hat{\Omega},f),
\ee
is a sum over all directions and polarizations,
 and  the detector response function
\be
F^{A}(\hat{\Omega},f)=
\textbf{D}(\hat{\Omega},f):\epsilon ^A(\hat{\Omega}).
\ee
$\mathcal R(f)$  is  determined by
the geometry of the interferometer,
and   transfers the incident stochastic RGW signal into the output signal.
A greater value of $\mathcal{R}(f)$ means
 a stronger ability  to transfer RGW into the output signal.
Formula  (\ref{transferGeneral})
applies to the Michelson, Sagnac, and  symmetrized Sagnac interferometers.
Using  $\textbf{D}_m $, $\textbf{D}_s$,  $\textbf{D}_{ss} $
of (\ref{response}), (\ref{sagnac response}) and (\ref{ss response}),
yields the transfer functions
$\mathcal{R}_m(f)$, $\mathcal{R}_s(f)$ and $\mathcal{R}_{ss}(f)$, respectively.
They are plotted in the top of Fig.~\ref{R+S}.
For the Michelson,    $R(f ) \simeq  0.3$    in low frequencies,
which is much greater than the
corresponding value for the Sagnac and symmetrized Sagnac,
so the Michelson  has a stronger  ability to
transfer  incident RGW   into the output signals.

\subsection{The sensitivity  of one interferometer and detection of RGW}

Including  the noise,
the total output signal of an interferometer   is  a sum
\be \label{total signal}
    s(t)=h_o(t)+n(t) ,
\ee
where $h_o(t) $ is   the output response of (\ref{strain})
and  $n(t)$   is a Gaussian noise signal with a zero mean $\langle n(t) \rangle=0$,
 uncorrelated to $h_o $.
Define
\be\label{noi}
\langle n(t)n(t')\rangle
=\frac12 \int^{+\infty}_{-\infty}df e^{i 2\pi f(t-t')}   S_n(f) ,
\ee
where $S_n(f)$ is the noise spectral density. It satisfies
\be \label{noise spectral}
\langle n^2(t)\rangle
= \frac12 \int^{+\infty}_{-\infty}df\ S_n(f)
=\int^\infty_0 df\ S_n(f) .
\ee
The noise in the frequency domain can be  equivalently specified by
\begin{equation}\label{noise emsemble}
\langle \tilde{n}(f) \rangle=0,
    ~~~~
\langle \tilde{n}^*( f)\tilde{n}(f') \rangle
=\frac{1}{2}\delta(f-f')S_n(f) .
\end{equation}
There are two major  kinds of noise \cite{Bender1998,Cornish}.
The first kind  is   called the optical-path noise,
which includes
shot noise, beam pointing instabilities, thermal vibrations, etc.
Among these, shot noise is the most important
and its  noise spectral density is given by  \cite{Cornish}
\be \label{Ss,Sa}
 S_s(f) =\frac{1.21\times10^{-22}{\rm m}^2\,{\rm Hz}^{-1}}{(5\times10^9{\rm m})^2}
  =4.84\times10^{-42}\ {\rm Hz}^{-1}.
\ee
The other kind of noise is the acceleration noise
with a spectral density
\be
S_a(f)=
\frac{9\times10^{-30}{\rm m}^2\,{\rm s}^{-4}\,{\rm Hz}^{-1}}
                {(5\times10^9{\rm m})^2 (2 \pi f)^4}
=
    \ 2.31\times10^{-40} \left(\frac{{\rm mHz}}{f}\right)^4\ {\rm Hz}^{-1} .
\ee
From these follow the noise spectral density of
the Michelson
\begin{equation}\label{noise spectral density 2}
S^m_n(f)=8S_a(f)
\left(1+\cos^2\left(\frac{f}{f_*}\right)\right)
+4S_s(f),
\end{equation}
the noise spectral density of the Sagnac
\begin{equation}\label{sns}
    S_n^s(f)=6S_s(f)+8\left(\sin^2\frac{3f}{2f_*}+2\sin^2\frac{f}{2f_*}\right)S_a(f),
\end{equation}
and the noise spectral density of the symmetrized Sagnac
\be\label{ss noise spectral density}
    S_n^{ss}(f)=\frac{2}{3}\left(1+2\cos\left(\frac{f}{f_*}\right)\right)^2
    \left(S_s(f)+4\sin^2\left(\frac{f}{2f_*}\right)S_a(f)\right).
\ee
These   are plotted   in the bottom of Fig.~\ref{R+S}.
It is seen that the noise of the  Michelson  is larger than
those of   Sagnac and    symmetrized Sagnac,
and the symmetrized Sagnac  has the least noise.
Higher symmetries in the optical path designs
of Sagnac and   symmetrized Sagnac cancel more noise.
The symmetric Sagnac has a transfer function
 several orders of magnitude lower than Michelson  of around $10^{-3}$\,Hz,
and  can be used to monitor
 noise level in practice \cite{TintoArmstrongEstabrook2000,Cornish}.
\begin{figure}[htbp]
\centering
  \includegraphics[width=0.6\columnwidth]{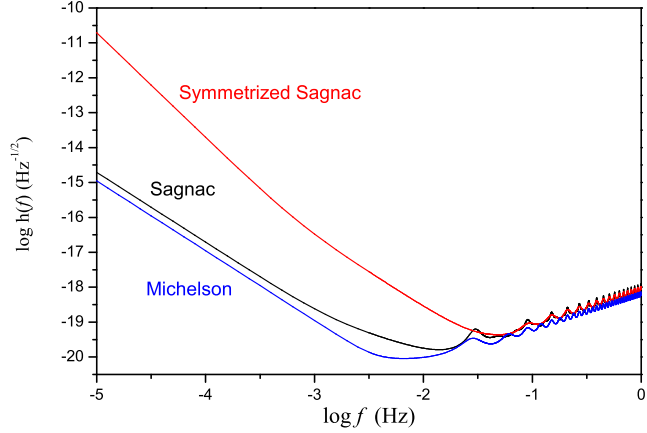}\\
  \caption{
  The sensitivity curves $\tilde{h}(f) $
  of Michelson, Sagnac and symmetrized Sagnac
  }\label{figure curves}
\end{figure}

\begin{figure}[htbp]
\centering
  \includegraphics[width=0.7\columnwidth]{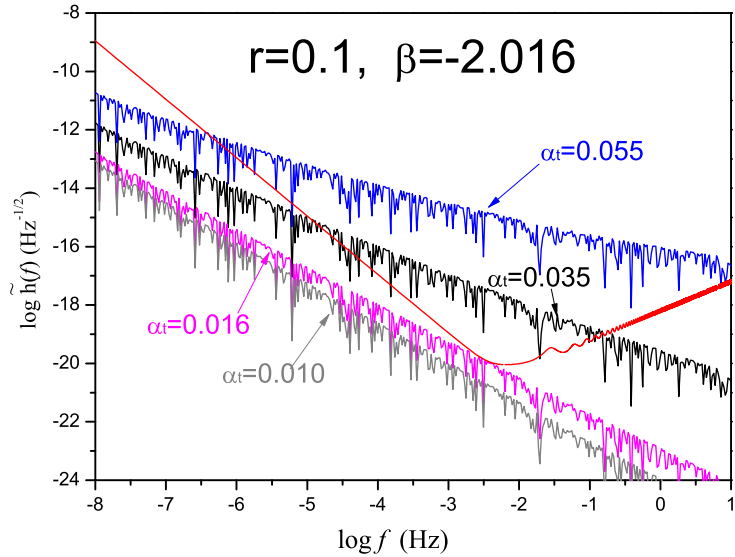}\\
  \caption{
 Comparison of $h_c(f) $ of RGW with the sensitivity
    $\tilde{h}(f)$ of a single interferometer.
       }\label{RGW-b-LISA}
\end{figure}

To detect RGW by signals from one interferometer,
one considers the auto-correlation of the total output signal
\bl\label{totaloutput}
\langle s^2(t)\rangle=&\langle h_o^2(t)\rangle
  +\langle n^2(t)\rangle
= \int ^\infty _0 df\  S (f),
\el
where   (\ref{sm ensemble average}) and (\ref{noise spectral}) are used,
and the total spectral density
\be\label{spectral density total}
S(f)\equiv   S_h(f) \mathcal{R}(f)+ S_n(f)   .
\ee
(\ref{totaloutput})  is equivalently written in the frequency domain
\be \label{ss'}
\langle \tilde s^*(f)\tilde s(f') \rangle
        =\frac12\delta(f-f') S(f)  .
\ee
Since both the RGW signal and noise occur in (\ref{spectral density total}),
SNR for a single interferometer which is denoted as SNR$_1$ can be naturally defined as
\be\label{SNRsingle}
\text{SNR}_1
\equiv
\frac{h_c(f)}{\tilde h(f)},
\ee
where $h_c(f)$ is related to $S_h(f)$ by (\ref{hspectrum}),
and  the sensitivity   is introduced by
\be \label{h response}
\tilde{h}(f) \equiv \sqrt{\frac{S_n(f)} {\mathcal{R}(f)} } ,
\ee
which reflects the detection capability of one interferometer.
A smaller  $\tilde{h}(f) $ indicates a better sensitivity,
which requires a lower $S_n$ and a greater ${\mathcal{R}} $.
Figure \ref{figure curves} shows
the   sensitivity curves  of three interferometers,
which is similar to the result of Ref.\cite{Cornish}.
It is seen that
the Michelson   has the best sensitivity level,
  $\tilde{h}(f) \sim 10^{-20} $ Hz$^{-1/2}$
 around $f_*= c/(2\pi L)\simeq10^{-2}$\,Hz
for the arm-length   $L=  5\times 10^9$\,m.
This is because the  transfer function $\mathcal R(f)$
of the Michelson is greatest,
giving rise to a lowest value for $\tilde h(f)$,
 even though its $S_n(f)$ is slightly higher than the other two.
Therefore, we shall use  the Michelson in the subsequent sections.
As a preliminary criterion,
a single interferometer will detect RGW   when SNR$_1$ $>1$,   i.e,
\be \label{hchtilde}
h_c(f) >  \tilde{h}(f).
\ee
This criterion was used
to  constrain  the RGW parameters
from the data of LIGO   S5 \cite{LIGOS5,Zhang2010}.
We plot $h_c(f) $ and  $\tilde{h}(f)$
in Fig.~\ref{RGW-b-LISA}  for a single  interferometer.
When the  data of space-borne interferometers are available
in the future,
(\ref{hchtilde}) will put a constraint on the   parameters.
The interferometer will be able to detect RGW of
$\alpha_t>0.016$ at fixed $r=0.1$ and $\beta=-2.016$.

\section{Estimation of RGW spectrum by one interferometer}

Now
we try to determine  the RGW spectrum
from the output signals of one   interferometer.
This is  a typical estimation problem of statistical signals,
which can be studied   by statistical  methods.
From the  view of statistics,
RGW and CMB  share  some similar properties,
both of them form a stochastic background in the Universe,
and can be modeled by a Gaussian random field \cite{Gorski1994,Oh1999,Hinshaw2003Apjs}.

The  time series  of the  output signal (\ref{total signal})
can be put into  the Fourier form  in    frequency domain
\[
\tilde s(f) =\tilde h_o(f)+  \tilde n(f).
\]
For practical computation,
the data set
can be divided into the following sample vector
\be\label{setdatas}
{\bf\tilde s}=[\tilde s(f_1),...,\tilde s(f_N)],
\ee
with $f_{i+1}-f_i=\Delta f$, $i=1,2,\cdots,N$,
where $N$ is a sufficiently large number.
Since both  $\tilde h_o(f_i)$ and $\tilde n(f_i)$,
are  Gaussian and independent,
$\tilde s(f_i)$ is a Gaussian random variable,
and ${\bf\tilde s}$   consists of $N$ statistically independent Gaussian data points,
having zero mean
\be\label{zeromean}
\langle\tilde s(f_i)\rangle=0 .
\ee
The covariance matrix  is  (\ref{ss'}),
which  is  written in the discrete form
\be\label{covarianceone}
\Sigma_{ij}
   = \delta_{ij}\frac1{2\Delta f}\, S(f_i),
                    ~~~~ i,j =1,2,\cdots,N
\ee
where the Dirac delta function  $\delta(f)$
has been  replaced by its discrete  form  \cite{Finn1992},
\be\label{deltaij3}
 \delta(f_i-f_j)
= \lim_{\Delta f\rightarrow0} \frac{1}{ \Delta f} \delta_{ij}.
\ee
The total spectral density in (\ref{spectral density total})
is also written in the discrete form
\be\label{Si}
S(f_i)\equiv  \l[S_h(f_i)\mathcal{R}(f_i)  +S_n(f_i)\r].
\ee
The inverse covariance matrix   is
\be\label{covariance matrix ij-1}
(\bm\Sigma^{-1} )_{ij}=
\frac{2\Delta f\delta_{ij}}{ S(f_i )  } \, ,
\ee
 depending on  the RGW signal $S_h $,
the noise $S_n $ and the transfer function $\mathcal R $
of the interferometer.
Note that here $\Sigma_{ij}$ is diagonal since
 $\tilde s(f_i)$ and $\tilde s(f_j)$ are independent for $i\neq j$.
Given the mean and the covariance,
the PDF of $\bf\tilde s$ is written as a multivariate Gaussian PDF \cite{Kay}
\be \label{gaussone}
f({\bf\tilde s})=
\frac{1}{(2\pi)^{\frac{N}{2}}\text{det}^{\frac12}[\bm{\Sigma}]}
\exp\l\{-\frac12 {\bf \tilde s}\ \bm{\Sigma}^{-1}\ {\bf \tilde s}^T\r\},
\ee
and the likelihood function is
\be\label{likeli2}
 \mathcal L
\equiv -  \ln f({\bf\tilde s})
= \frac12 \ln \text{det} [\bm{\Sigma}]
 +\frac12 { {\bf \tilde s} }\ \bm{\Sigma}^{-1}{\bf \tilde s} ^T \, ,
\ee
 (dropping an irrelevant additive constant $\frac12 N \ln 2\pi$).
Once the PDF  is chosen,
 an estimator  of the spectrum  is a specification
to give the value $S_h$ for the given data set $\bm \tilde s$.
For this,  we shall adopt the ML method.
In general,
$\mathcal L$ can be expanded in a neighborhood of  some spectrum $\bar S_h(f)$,
\bl
 \mathcal L = & \bar{ \mathcal L}
 +
 \sum_{k=1}^N\frac{\partial  \mathcal L}{\partial  S_{h}(f_k)}(S_{h}(f_k)-\bar S_{h}(f_k))
    \nn \\
&   +\frac12
\sum_{k,\,l=1}^N \frac{\partial  ^2 \mathcal L}{ \partial  S_{h}(f_k) \partial S_h(f_l) }
  \l( S_{h}(f_k) - \bar S_{h}(f_k) \r) \l(S_h(f_l)- \bar S_h(f_l)\r).
\el
We  look for  the  most  likely  spectrum $\bar S_h(f)$
at which  $  \mathcal L  $  is minimized
\be
\l.\frac{\delta  \mathcal L}{\delta  S_{h} }\r|_{\bar S_h}  =0.
\ee
Taking the derivative of Eq.~(\ref{likeli2}) with respect to $S_h(f_i)$,
and using the relations
\be\label{matrix relation3}
\frac{\partial\ln\text{det}[\bm\Sigma ]}{\partial  S_{h}(f) }
=\text{tr}\l(\bm\Sigma^{-1}
\frac{\partial\bm\Sigma }{\partial   S_{h}(f)} \r),
~~~~
\frac{\partial\bm\Sigma^{-1}  }
{\partial   S_{h}(f)}
=-\bm\Sigma^{-1}
\frac{\partial\bm\Sigma }{\partial   S_{h}(f) }
\bm\Sigma^{-1},
\ee
one produces the first order   derivative \cite{Kay}
\bl\label{PartialLikelyhood3}
\frac{\partial\mathcal L}
{\partial S_{h}(f_i)}
=&\frac12\text{tr}\l(\bm\Sigma^{-1}
\frac{\partial\bm\Sigma}{\partial S_{h}(f_i)}\r)
-\frac12\bm{\tilde s}\
\bm\Sigma^{-1}
\frac{\partial\bm\Sigma}{\partial S_{h}(f_i)}
\bm\Sigma^{-1}\ \bm{ \tilde s}^T \, ,
\el
where  the zero mean of $\bm s$ gives no contribution to  (\ref{PartialLikelyhood3}).
From Eq.~(\ref{covarianceone}),  one calculates
\be\label{pCorSh}
\frac{\partial\Sigma_{kl}}{\partial S_h(f_i)}
=
\frac12\frac1{\Delta f}\delta_{kl}\delta_{ki}\mathcal{R}(f_k),
 ~~~~ k,l,i =1,2,\cdots,N ,
\ee
where
\be\label{ShiPShj}
\frac{\partial S_h(f_i)}{\partial S_h(f_j)}=\delta_{ij},
~~~~
i,j=1,\cdots,N .
\ee
has been  used.
Substituting (\ref{covarianceone}) and (\ref{pCorSh}) into (\ref{PartialLikelyhood3})
     leads to
\bl\label{PartialLikelyhood4od}
\frac{\partial\mathcal L}
{\partial S_{h}(f_i)}
=&\frac{\mathcal{R}(f_i)}{2 [S(f_i)]^2}
\l[   S_h(f_i)\mathcal{R}(f_i) +S_n(f_i) -2\tilde s_i^2\Delta f  \r] .
\el
Setting this to zero,
one obtains the ML estimator of the RGW spectrum
\be\label{ShEstimator}
    S_h(f_i)
=\frac{2\Delta f}{\mathcal{R}(f_i)} \tilde s_i^2
-\frac{S_n(f_i)}{\mathcal{R}(f_i)},
 ~~~~ i =1,2,\cdots,N .
\ee
However, in practice, our knowledge is not sufficient
for the spectrum $S_n(f_i)$
due to noise that is actually inherent in the data
 \cite{Flanagan1993,TintoArmstrongEstabrook2000},
so that the formula (\ref{ShEstimator}) for a single
is not effective to estimate the RGW spectrum
when the noise is dominantly large.
In the following we turn to two detectors for estimation of RGW spectrum.

\section{ Cross-correlation of a pair of  interferometers}

\subsection{The overlapping function of a pair  of
case for LISA}

\begin{figure}[htbp]
\centering
  \includegraphics[width=0.6\columnwidth]{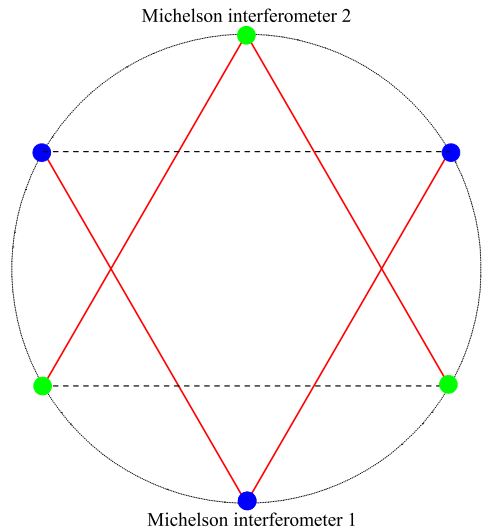}\\
  \caption{A pair of interferometers in  two triangles for  config. 2
  }\label{LISAcorrelate12}
\end{figure}

We  present a pair of Michelson  interferometers in space
which has been studied in Refs.\cite{Cornish & Larson,Cornish}.
Based on
the three spacecraft forming  on an equilateral  triangle
 in Figure \ref{figure structure},
we consider two  configurations of a pair.
Config. 1 consists of the three spacecraft in one triangle,
but with each spacecraft  carrying two sets of independent detection equipments.
Then two independent Michelson interferometers form \cite{Cutler1998}:
the first having  the point 1 as the vertex,
and the second having  the point 2 as the vertex,
differing from the first by a rotation of $120^\circ$,
as in Figure \ref{figure structure}.
This configuration is economically favored,
however, a possible problem is that
the equipment on one craft may have dependent noise.
For simplicity we assume that the noises are independent
by better  setup in the design.
Config. 2 consists of  two   triangles,
equipped with six spacecraft, forming two   interferometers
        in space \cite{Cornish & Larson}.
The   second triangle is rotated $180^\circ$
from the first,  as in Figure \ref{LISAcorrelate12}.
This configuration can ensure independent noise
since the six crafts are located far away
from each other in space.
For both configurations,
we assume that the noises from the two interferometers are
independent of  each other, and independent of RGW.
GW signals are  correlated in the two interferometers.
By  cross-correlating the output data of the pair,
the detection capability   will be enhanced.
Consider  the output signals from a pair of two interferometers
\bl
s_1(t)&=h_1(t)+n_1(t),\label{output individually1} \\
s_2(t)&=h_2(t)+n_2(t),\label{output individually2}
\el
each is similar to Eq.~(\ref{total signal}).
It is assumed that
\be\label{noise uncorrelated}
\langle h_i(t)n_j(t)\rangle=0,~~~~~~
\langle n_1(t)n_2(t)\rangle=0,
\ee
and
\be\label{moisesp}
\langle n^2_1(t)\rangle=\int_0^\infty df S_{n1}(f),~~~~~~
\langle n^2_2(t)\rangle=\int_0^\infty df S_{n2}(f),
\ee
where $S_{n1} $, $S_{n2} $ are the noise spectral densities
of  the two interferometer as in (\ref{noise spectral}),
and specified as in Eq.~(\ref{noise spectral density 2}) for the  Michelson.
When the two interferometers  are identical,
one can take $S_{n1} \simeq S_{n2} $.

\begin{figure}[htbp]
\centering
  \includegraphics[width=0.6\columnwidth]{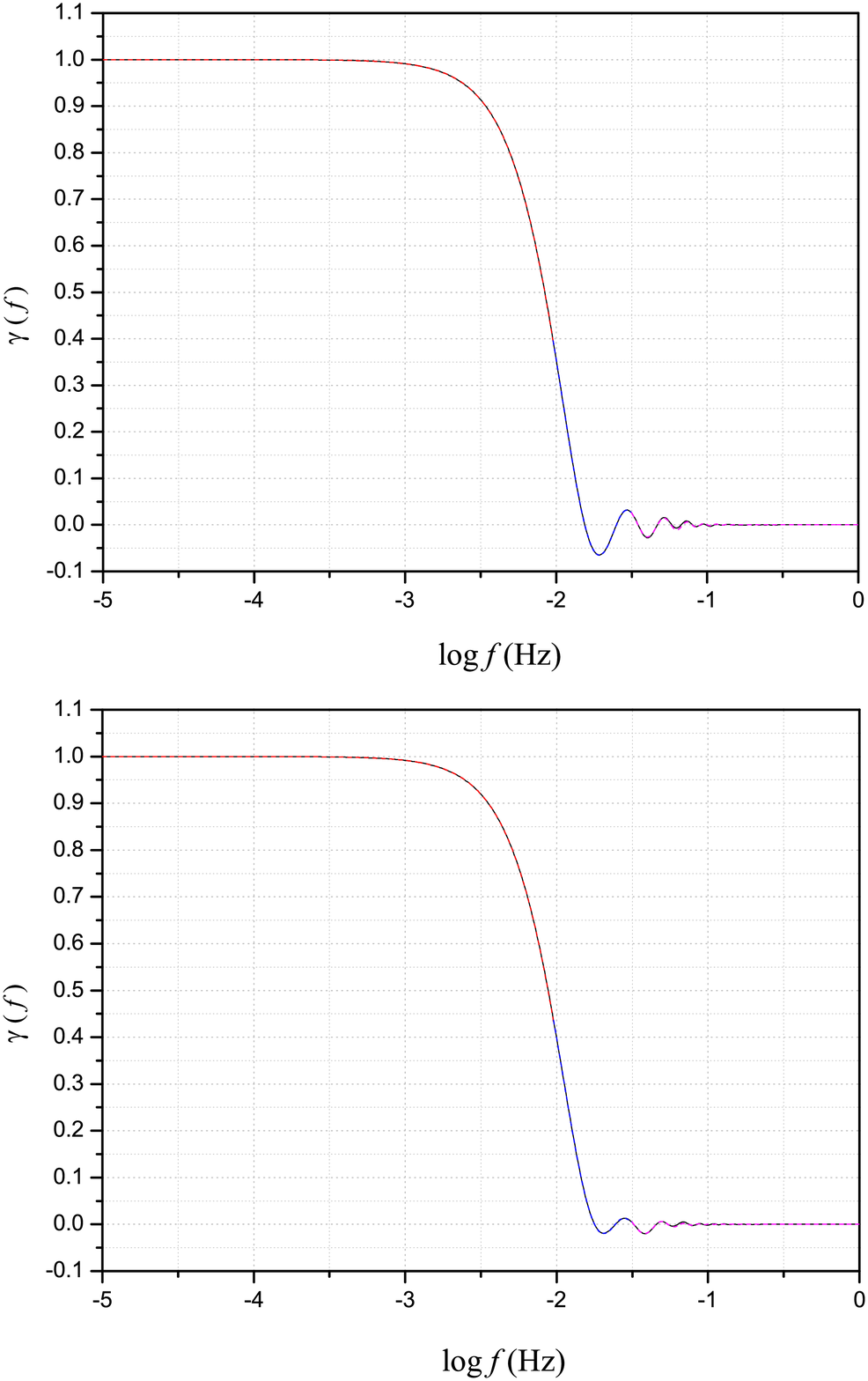}\\
  \caption{
  The overlap reduction function for a pair.
  Top: Config.1;
  Bottom: Config.2.
 {\it Solid lines} are plotted numerically
    and {\it dotted lines} by the fitted formulae (\ref{overlappingapprox1}) and (\ref{overlappingapprox2}).
     }\label{2overlapping}
\end{figure}

Using the output response of (\ref{sm expansion}) for each  interferometer
and formula (\ref{h2 emsemble}),
the ensemble average of the correlation  of two output responses is
\bl
\langle h_1(t)h_2(t')\rangle
=&\frac12\int_{-\infty}^{+\infty}df S_h(f)\int\frac{d\hat{\Omega}}{4\pi}\sum_{A=\times, +}
F_1^{A*}(\hat{\Omega},f)F_2^{A}(\hat{\Omega},f)
e^{- i 2\pi f\hat\Omega\cdot(\mathbf r_1-\mathbf r_2)} e^{i2\pi f(t-t')} \nonumber
\\
=&\int_0^\infty d f S_h(f)\mathcal R_{12}(f)e^{i2\pi f(t-t')} ,
\el
similarly, by (\ref{hij}) and (\ref{h2 emsemble}),
\bl\label{ensemble average of s1s2}
\langle\tilde h_1(f)\tilde h_2(f')\rangle
=&\frac{1}{2}\delta (f-f')  S_h(f)\mathcal R_{12}(f),
\el
where the transfer function $\mathcal R_{12}(f)$ is
\be\label{correlated transfer function}
\mathcal{R}_{12}(f)\equiv\int\frac{d \hat{\Omega} }{4\pi}\sum_{A=\times, +}
F_1^{A*}(\hat{\Omega},f)F_2^{A}(\hat{\Omega},f)e^{- i 2\pi f\hat\Omega\cdot(\mathbf r_1-\mathbf r_2)},
\ee
where   $\mathbf r_1$ and $\mathbf r_2$ represent
the respective positions of the vertices of the two interferometers.
A higher value of $\mathcal{R}_{12}(f)$ means a better capability to transfer incoming RGW into signals from the detector.
\begin{figure}[htbp]
\centering
  \includegraphics[width=0.8\columnwidth]{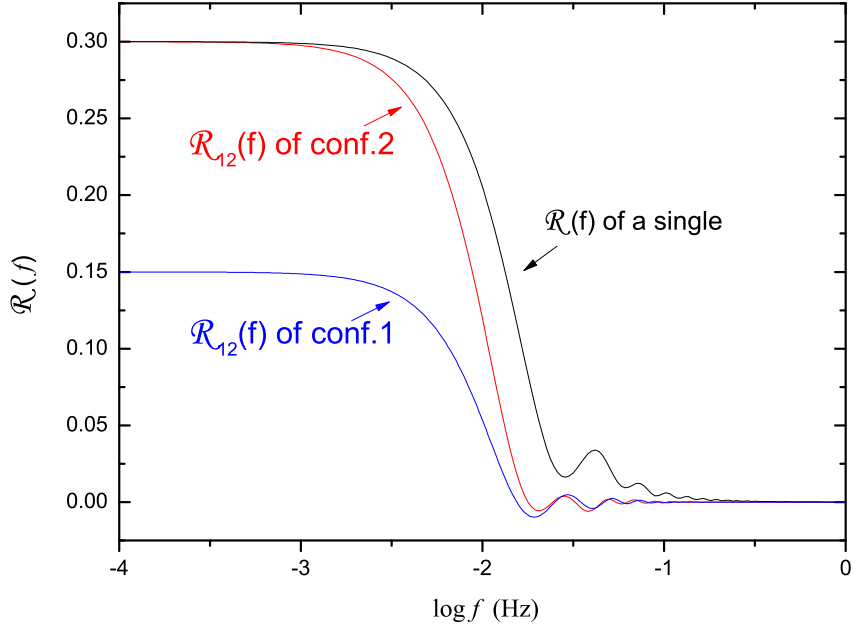}\\
  \caption{
  A comparison between the transfer function of a single
  and a pair.
   }\label{R121compare}
\end{figure}
In Fig. \ref{R121compare}
we plot the transfer function of a single  and of  a pair
with conf.1 and conf.2.
One introduces the overlapping reduction function $\gamma(f)$
              \cite{Flanagan1993,Allen & Romano}
by normalizing $\mathcal R_{12}(f)$ as the following
\be\label{overlapping}
\gamma(f) \equiv  \frac5{\sin^2\beta_0}\mathcal R_{12}(f)
= \frac{20}{3}\mathcal R_{12}(f).
~~~~{\rm for ~ conf.1}
\ee
\be\label{overlapping2}
\gamma(f) \equiv \frac5{ 2 \sin^2\beta_0}\mathcal R_{12}(f)
=  \frac{10}{3}    \mathcal R_{12}(f).
~~~~{\rm for ~ conf.2}
\ee
where $\beta_0=\pi/3$ is the angle between  arms of one interferometer,
 $\sin^2\beta_0=3/4$.
Clearly, $\gamma (f)$  depends on the geometry of the pair,
and transfers the incident RGW from all the directions
 (by integration over angle $\hat{\Omega}$)
  into the output signal.
We  compute $\gamma(f)$ numerically and plot it
in Fig.~\ref{2overlapping} for  config.1 {\it(top)}
and config.2 {\it(bottom)}.
At high frequencies,  $\gamma(f) $  oscillates around zero.
At low frequencies,   $\gamma(f) \rightarrow 1$
 for both  configurations.
To a high accuracy, it can be  fitted by the following formula
\be \label{overlappingapprox1}
\gamma(f)=\left\{
           \begin{array}{ll}
              1-0.811508\l(\frac{f}{f_*}\r)^2
              +0.241292\l(\frac{f}{f_*}\r)^4
              -0.0374118\l(\frac{f}{f_*}\r)^6,
             &
             f< f_*
              \\
              0.43636
              +2.12337\l(\frac{f}{f_*}\r)
              -4.00143\l(\frac{f}{f_*}\r)^2
                +2.37321\l(\frac{f}{f_*}\r)^3
            \\
              \hspace{4.4cm}
              -0.588745\l(\frac{f}{f_*}\r)^4
              +0.0528759\l(\frac{f}{f_*}\r)^5,
             &
              f_*\le f < 3.3f_*
              \\
              0.000311254-0.11762
              e^{-0.37176f/f_*}\bigg[0.0849126
              \\
              \hspace{2.6cm}
             -\sin\bigg(2.89649
             -2.40829\l(\frac{f}{f_*}\r)e^{0.00978685f/f_*}\bigg)\bigg],
   &
     3.3f_*\le f
    \end{array}  \right .
\ee
for the config. 1,
and
\be  \label{overlappingapprox2}
\gamma(f)=\left\{
           \begin{array}{ll}
              1-\frac{383}{504}\l(\frac{f}{f_*}\r)^2
              +\frac{893}{3888}\l(\frac{f}{f_*}\r)^4
              -\frac{5414989}{143700480}\l(\frac{f}{f_*}\r)^6,
             &
             f< f_*
              \\
              0.629524
              +1.52435\l(\frac{f}{f_*}\r)
              -3.23303\l(\frac{f}{f_*}\r)^2
                +1.96633\l(\frac{f}{f_*}\r)^3
            \\
              \hspace{4.4cm}
              -0.496608\l(\frac{f}{f_*}\r)^4
              +0.0454382\l(\frac{f}{f_*}\r)^5,
             &
              f_*\le f < 3.3f_*
              \\
              0.000190192-0.157835
              e^{-0.570049f/f_*}\bigg[0.264226

              \\
              \hspace{2.6cm}
             -\sin\bigg( 1.66131
             -2.01502\l(\frac{f}{f_*}\r)e^{0.0349379f/f_*}\bigg)\bigg],
   &
     3.3f_*\le f
    \end{array}  \right .
\ee
for  config.2.
 $\gamma(f)$ of   configs. 1 and 2 look the same
except around $f \simeq (2 \sim 8)\times 10^{-2}$\,Hz.
Formula (\ref{overlappingapprox2}) at $f< f_*$ agrees with
that in Ref.\cite{Cornish & Larson},
which does not list the expression for the part $f>  f_*$.
See also Ref.\cite{UngarelliVecchio2001}.
In the following we shall mainly use config. 2 for demonstration.

\subsection{SNR for a pair case for LISA}

To suppress the noise of a pair,
one  defines the   cross-correlated, integrated  signal
of $ s_1(t)$ and $s_2(t')$ as the following \cite{Allen & Romano}
\bl\label{correlation signal}
C&\equiv\int_{-T/2}^{T/2}dt\int_{-\infty}^{+\infty}
dt' s_1(t)s_2(t') Q(t-t')\nonumber\\
& = \int_{-\infty}^\infty df\int_{-\infty}^\infty df'
  \delta_T(f-f') \tilde s_1^{\, *}(f)   \tilde s_2(f')\tilde Q(f'),
\el
where $T$ is the observation time,
$\tilde s(f)=\tilde s(-f)$ is the Fourier transform of $s(t)$
and $\tilde s^{\, *}  _1(f)=\tilde s_1(-f)$,
 $Q(t-t')$ is a filter function to be determined by maximizing SNR$_{12}$ (SNR for the pair),
its Fourier transform is $\tilde Q(f)=\tilde Q(-f)$,
and
\be\label{dedeltaT}
\delta_T(f) \equiv
    \int_{-T/2}^{T/2}dt\, e^{-i 2\pi f t}=\frac{\sin(\pi f T)}{\pi f} ,
\ee
 is the finite-time Dirac delta function.
For a finite  $T$, one has $\delta_T(0)=T$,
and in the limit $T\rightarrow\infty$,
$\delta_T(f)$ reduces to the Dirac delta  function $\delta(f)$.
Given the   frequency band of  $(10^{-4}- 10^{-1})$ Hz,
one can take  the length of each segment, say,
$T \simeq 3$\,hours$\sim10^4$\,s.
When $T$ is large enough,
$\delta_T(\Delta f)$ is sharply peaked
over a narrow region of width   $\sim 1/T$.
Thus,  in the integration  (\ref {correlation signal}),
the product of  $\tilde s^*(f) \tilde s(f')$  contributes
only in the region of $|f-f'|< 1/T \sim  10^{-4}$ Hz.
The frequency band    contains $\sim 10^3$ of these regions.
By the central limit theorem,
$C$ is well-approximated  by a Gaussian random variable.

\begin{figure}[htbp]
\centering
  \includegraphics[width=0.6\columnwidth]{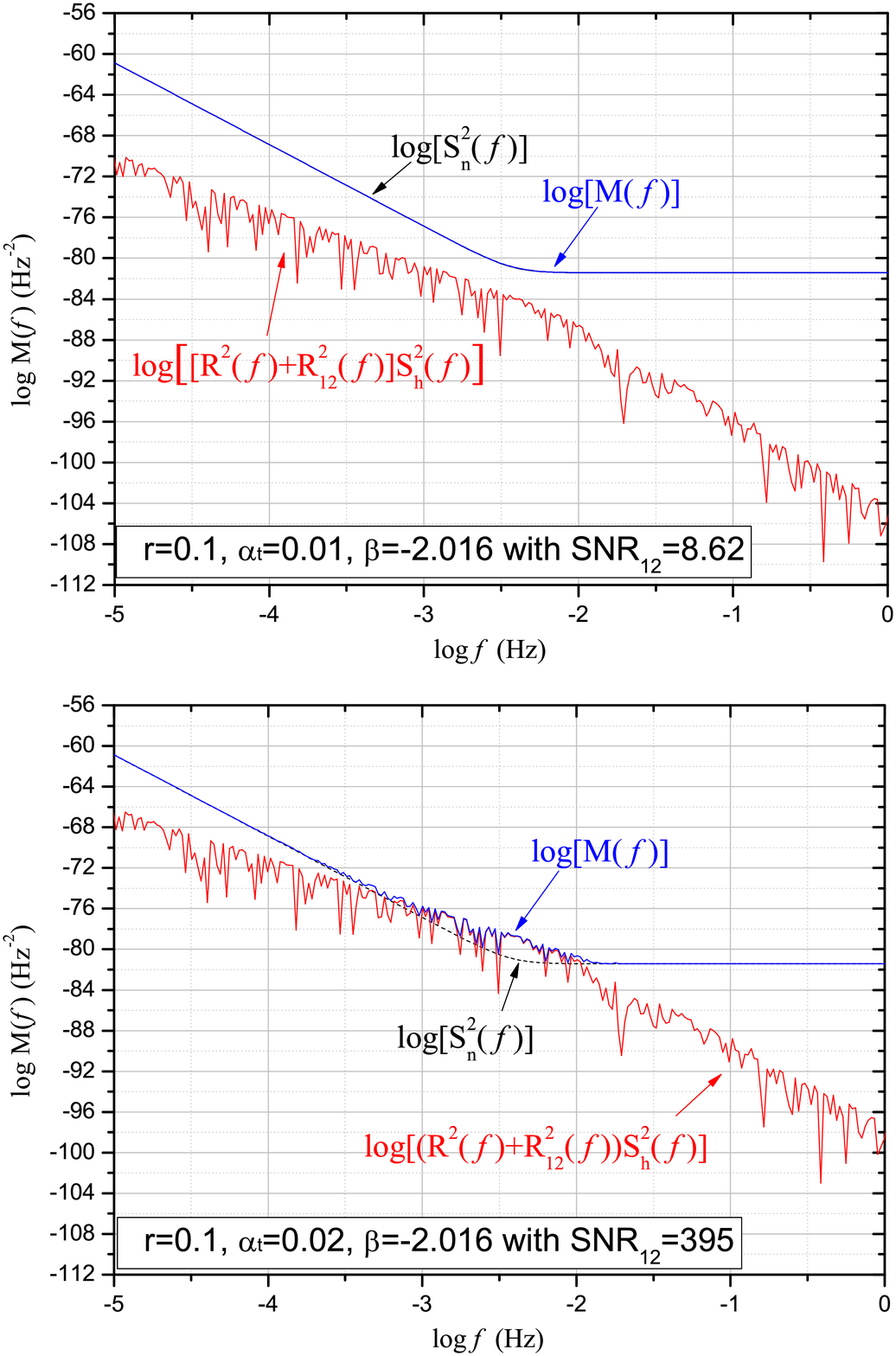}\\
  \caption{  \label{2MS}    $M$, $S^2_n$,
$\l( \mathcal R^2 + \mathcal R_{12}^2  \r) S_h^2$.
Top:  for SNR$_{12}=8.62$;
Bottom: for SNR$_{12}=395$.}
\end{figure}

In actual computations,
the cross-correlated signal $C$ can be expressed
either  in  the time domain, or  equivalently, the frequency domain.
In the following we shall use the frequency domain.
By Eqs. (\ref{noise uncorrelated}) and (\ref{ensemble average of s1s2}),
the mean of $C$   is
\bl\label{average correlation}
\mu
= \langle C\rangle
=&\int_{-T/2}^{T/2} df\int_{-\infty}^\infty df'\delta_T(f-f')\langle
\tilde s_1^*(f)\tilde s_2(f')\rangle
\tilde Q(f')\nonumber\\
=   &\frac {3T} {10} \int_{0}^\infty df S_h(f)\gamma(f)\tilde Q(f).
\el
Notice that the mean $\mu$ is non-zero,
in contrast to  (\ref{zeromean}) for one interferometer.
Furthermore,
the noise terms disappear in the above
since they are removed by cross-correlation,
and only the RGW signal accumulates with the observation time $T$.
This  feature of a pair is the advantage over  a single case.
A greater value of $\mu$ is desired for RGW detection.
The covariance $C$  is
\bl \label{sigma2}
\sigma^2
=&\langle C^2\rangle-\langle C\rangle^2
\\
=&\int_{-\infty}^{\infty}df\int_{-\infty}^\infty df'
 \int_{-\infty}^{\infty}dk\int_{-\infty}^{\infty}dk'
 \delta_T(f-f')\delta_T(k-k')\tilde Q(f')\tilde Q^*(k')
\nonumber\\
&
\times\l\langle \l[\tilde h_1^*(f)+\tilde n_1^*(f)\r]
\l[\tilde h_2(f')+\tilde n_2(f')\r]
\l[\tilde h_1(k)+\tilde n_1(k)\r]
\l[\tilde h^*_2(k')+\tilde n^*_2(k')\r] \r\rangle
\nn\\
&
-\bigg(\int_{-\infty}^\infty df\int_{-\infty}^\infty df'
 \delta_T(f-f')\frac12\delta(f-f')S_h(f')\mathcal R_{12}(f)
\tilde Q(f')\bigg)^2 .
\el
Using  the ``factorization"  property \cite{Allen & Romano}
\[
\langle x_1x_2x_3x_4\rangle
=\langle x_1x_2\rangle \langle x_3x_4\rangle
+\langle x_1 x_3\rangle \langle x_2x_4\rangle
+\langle x_1x_4\rangle \langle x_2 x_3\rangle \, ,
\]
valid for Gaussian random variables $x_1,x_2,x_3,x_4$,  each having zero mean,
and using Eqs.  (\ref{Sh normalization2}),
(\ref{noise uncorrelated}) and (\ref{ensemble average of s1s2}),
one obtains
\bl\label{variance correlation0}
\sigma^2
=&\frac 14\int_{-\infty}^\infty df\int_{-\infty}^\infty
df' \delta^2_T(f-f')\l|\tilde Q(f')\r|^2
\Big[
S_{1n}(f)S_{2n}(f')
+\mathcal R(f')S_h(f')S_{1n}(f)\nonumber
\\
&+\mathcal R(f)S_h(f)S_{2n}(f')
+\mathcal R(f)S_h(f)\mathcal R(f')S_h(f')
+   \mathcal R_{12}(f)S_h(f)\mathcal R_{12}(f')S_h(f') \Big] ,
\el
where $ \mathcal R(f)$ is the transfer function for a single case in (\ref{transferGeneral}),
and $\mathcal R_{12}(f)$ is the transfer function for
a pair  in  (\ref {correlated transfer function}).
For $T$   sufficiently long,
one $\delta_T(f-f')$ can be set to be
the  Dirac function $\delta(f-f')$, yielding
\bl\label{variance correlation1}
\sigma^2
= \frac{T}{2}\int_{-\infty}^\infty df  \, |\tilde Q(f)|^2  \,  M(f) \, ,
\el
where the function
\bl \label{MS12}
M(f)\equiv
   S_{1n} (f) S_{2n} (f)
    +  \mathcal R(f) \Big[ S_{1n}(f) +  S_{2n}(f) \Big]S_h(f)
    + \Big[ \mathcal R^2(f) +  \mathcal R_{12}^2(f) \Big] S_h^2(f) ,
\el
which reduces to
\be \label{MS}
M(f) \simeq    S^2_n(f)  +2\mathcal R(f)S_h(f)S_n(f)
          + \l[ \mathcal R^2(f) + \mathcal R_{12}^2(f)  \r] S_h^2(f)
\ee
when  $ S_{1n}   \simeq S_{2n}  =S_n$.
We plot the functions $M $, $S^2_n $, and
$\l( \mathcal R^2  +  \mathcal R_{12}^2 \r) S_h^2 $ of (\ref {MS})
for different values of parameters  in Figs.\ref{2MS}
with SNR$_{12}$ given by (\ref{SNR max correlation}).
 $M$ is dominated  by $S^2_n$ at reasonable  values of  SNR$_{12}$,
so that one can take
$ M (f) \simeq  S_{1n} (f) S_{2n} (f)$
as a good approximation.

\begin{figure}[htbp]
\centering
  \includegraphics[width=0.6\columnwidth]{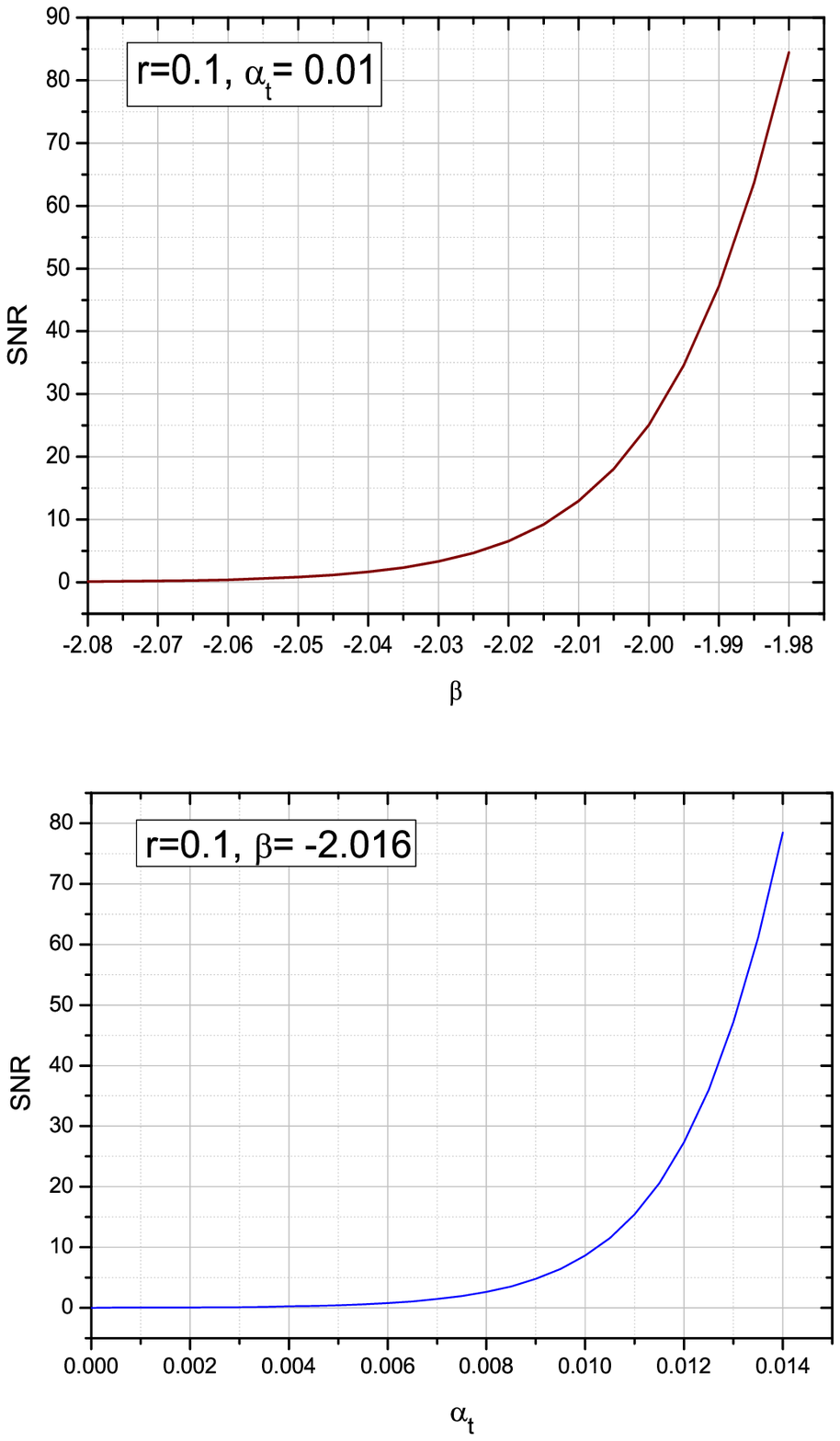}  \\
\caption{ SNR$_{12}$ changes with $\beta$ {\it(top)} and  $\alpha_t$ {\it (bottom)}.   }\label{2SNR}
\end{figure}

The  SNR  of the pair is defined as \cite{Allen & Romano}
\be
\text{SNR}_{12} =\frac{\mu}{\sigma} = \frac { 3 \sqrt {T}} {10}
\frac{\int_0^\infty df S_h(f)\gamma(f)\tilde Q(f)}{\left[\int_0^\infty df \,
  |\tilde Q(f)|^2 M(f)\right]^{1/2}},
\ee
which  describes the detection capability of a pair.
To maximize  $\text{SNR}_{12}$,
one chooses  the  filter function  \cite{Allen & Romano}
\be\label{Qf}
\tilde Q(f)= \frac{S_h(f)\gamma(f)}{M(f)},
\ee
for which the mean is
\be \label{meanmu}
\mu =  \frac {  3 T } {10}  \int_{0}^\infty df
                \frac{S^2_h(f)\, \gamma^2(f)}{M(f)} ,
\ee
the covariance is
\be  \label{mutosigma}
 \sigma^2 = \frac {T} {2} \int_{0}^\infty df
                \frac{S^2_h(f)\, \gamma^2(f)}{M(f)}
          =   \frac{5 }{3}  \mu \, ,
\ee
and
\be   \label{SNR max correlation}
\text{SNR}_{12}  =\frac{3 \sqrt{2T}}{10}
\left[\int_0^\infty df\frac{\gamma^2(f) S_h^2(f)}{M(f)}\right]^{1/2}.
\ee
$\text{SNR}_{12} \propto r$ for large noise,
and its dependences on $\beta$ and $\alpha_t $ are shown in  Fig.~\ref{2SNR}.
When the noise   is   dominant,
Eq.~(\ref{meanmu})  becomes
\be \label{NLmu}
\mu =  \frac { 3  T} {10}   \int_{0}^\infty df
                \frac{ S^2_h(f) \, \gamma^2(f)}{S_{1n}(f) S_{2n}(f)} \, ,
\ee
(\ref{SNR max correlation})  becomes
\bl
\text{SNR}_{12}& =\frac{ 3 \sqrt{2T}}{10}  \left[\int_0^\infty df
\frac{\gamma^2(f) S_h^2(f)}{S_{1n}(f) S_{2n}(f)}\right]^{1/2}.
\label{SNR12}
\el
Formula (\ref{SNR12}) is  similar to that of the ground-based  LIGO
\cite{Allen & Romano,Flanagan1993,Cornish & Larson}.
Clearly,
the dependence of SNR$_{12}$ on $r$,  $\beta$ and $\alpha_t$ is implicitly
contained in $S_h(f) $,
and SNR$_{12}\propto r$ for large noise.
There is a growing factor $\sqrt T$ of  $\text{SNR}_{12}$ in (\ref {SNR12}),
because the noise gets suppressed  by cross-correlation
and only the RGW signals accumulate with time.

To  demonstrate the capability of a pair of space interferometers
to detect RGW,
we compute  the values of SNR$_{12}$ using Eq.~(\ref{SNR max correlation}).
The result is in Table \ref{SNR LISA-corr and LIGO},
with an observation  time $T=1$ year and  $r=0.1$.
For comparison, we have also attached the result for
the pair of ground-based LIGO S6  \cite{LIGO and VirgoS6,LIGOnoise}
and LIGO O1 and Advanced LIGO as well \cite{GW150914},
for which we use the formulae of $\text{SNR}_{12}$
and  $\gamma(f)$ in Ref.\cite{Allen & Romano}.
It is seen that SNR$_{12}$  of LISA is higher than that of Advanced LIGO
by 4 orders of magnitude for the default $(\alpha_t=0,\beta=-2)$,
and  by 5 orders of magnitude for the observed-inferred  $(\alpha_t=0,\beta=-2.016)$.
Therefore,   LISA  will  have a much stronger capability  than LIGO to  detect  RGWs.
\begin{table}[htbp]
\Large
\caption{SNR$_{12}$ for a pair case for LISA  and for a pair
case for LIGO
with $r=0.1$.
}
\begin{center}
\label{SNR LISA-corr and LIGO}
\begin{tabular}{|c|c |c| c| c| c| c|}
 \hline
$\alpha_t,\beta$&
-0.005, -2.05&
0, -2.016&
0, -2&
0.005, -1.95&
0.01, -1.9\\
 \hline\hline
LIGO S6 &
$1.6\times10^{-13}$ &
$5.0\times10^{-10}$&
$2.2\times10^{-9}$ &
$3.1\times10^{-5}$ &
$4.2\times10^{-1}$\\\hline
LIGO O1 &
$3.0\times10^{-11}$ &
$7.3\times10^{-8}$&
$3.0\times10^{-7}$ &
$3.0\times10^{-3}$ &
$3.1\times10^{1}$\\\hline
Advanced LIGO &
$3.3\times10^{-10}$ &
$7.3\times10^{-7}$&
$2.9\times10^{-6}$ &
$2.7\times10^{-2}$ &
$2.5\times10^{2}$ \\\hline
A pair case for LISA &
$1.1\times10^{-4}$ &
$2.2\times10^{-2}$&
$6.8\times10^{-2}$ &
$3.8\times10^{1}$ &
$4.9\times10^{2}$ \\\hline
\end{tabular}
\end{center}
\end{table}

\subsection{The sensitivity of a pair compared with a single}

\begin{figure}[htbp]
\centering
  \includegraphics[width=0.6\columnwidth]{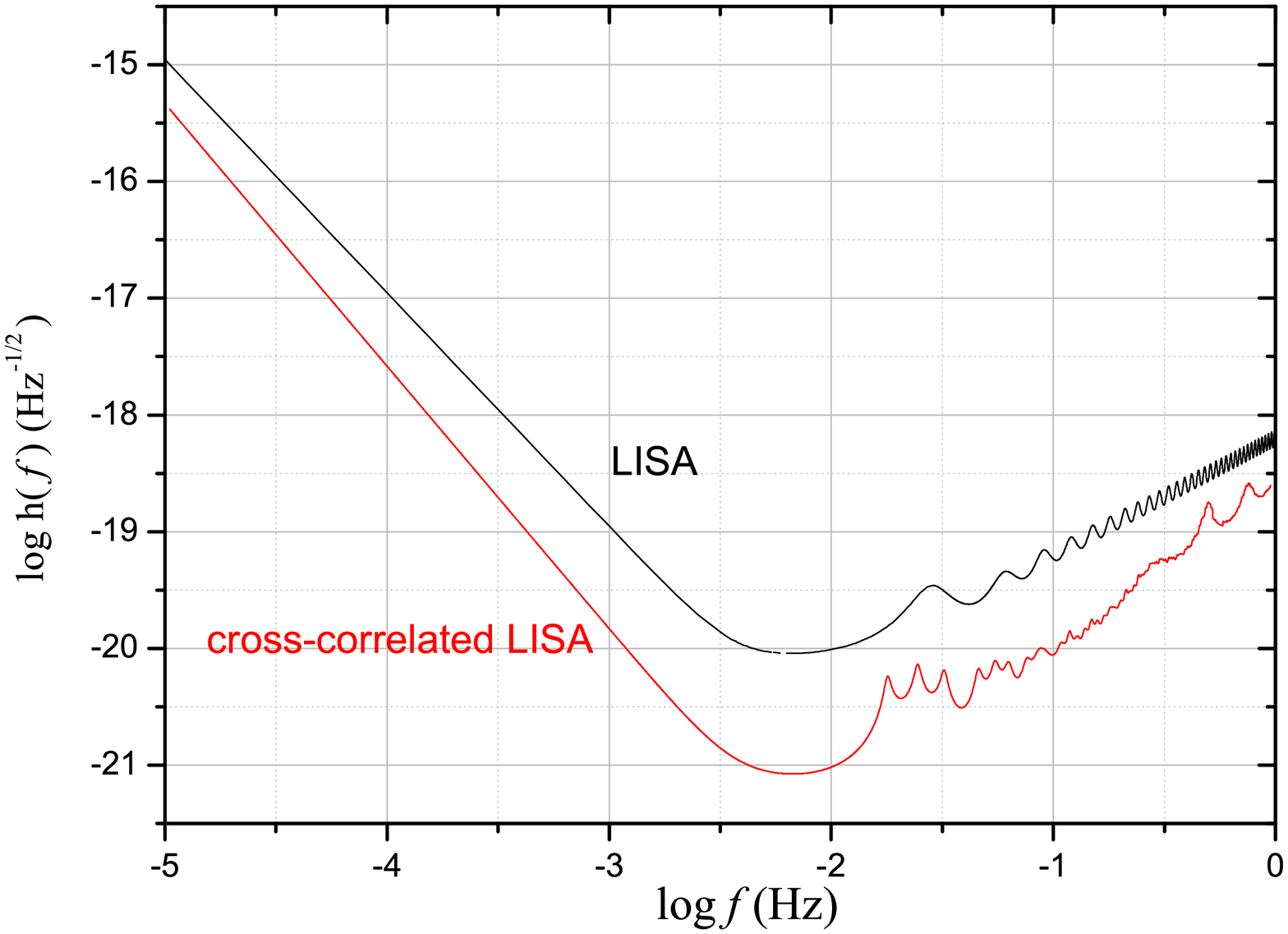}\\
  \caption{
 The sensitivity curves of a single   and a pair.
      }\label{figure h correlation}
\end{figure}

To  describe   the sensitivity of a  pair,
we can extend the expression of (\ref{SNR12})
and allow SNR$_{12}$ to vary with frequency.
Consider  the averaged  SNR$_{12}$ over a frequency band
of width $\Delta f$, centered at $f$  as the following \cite{Cornish}
\bl
\text{SNR}_{12}(f)
&\simeq   \sqrt{T}   \left[\int_{f-\Delta f/2}^{f+\Delta f/2} df
\frac{\mathcal R_{12} ^2(f) S_h^2(f)}{\left(S_n(f)\right)^2}\right]^{1/2}
\nn\\
&\simeq  \sqrt{T}  \sqrt{\Delta f} \,   S_h(f) \,
\overline{\l( \frac{\mathcal R_{12} ^2(f)}{\left(S_n(f)\right)^2} \r)}^{\,1/2} \, .
\el
In analogy to (\ref{h response}),
we define the effective sensitivity  of the   pair
\be\label{sensi}
\tilde h_{12}(f)  \equiv \sqrt{ \frac{S_h(f)}{\text{SNR}_{12}(f)  } }
  = \frac{1}{(T\Delta f)^{1/4}}
\overline{\l(
    \frac{\mathcal R_{12} ^2(f)}{\left(S_n(f)\right)^2} \r)} ^{\  -1/4},
\ee
which depends on  $T$ and frequency resolution $\Delta f$,
in contrast to that of a single  in (\ref{h response}).
A  longer $T$  increases the sensitivity of (\ref{sensi}).
The sensitivities  of a single
and  a pair are plotted in Fig.~\ref{figure h correlation},
where   $\Delta f=f/10$ and $T=1$ year are taken.
Clearly, a pair has a better sensitivity than a single
by $\sim 100$ times around $f\sim 10^{-2}$ Hz.

\subsection{ Constraints on the RGW parameters by a pair}

By (\ref{SNR12}),
a constraint on SNR$_{12}$ will transfer into
a constraint  on  ($r$, $\beta$, $\alpha_t$).
One such a constraint on SNR$_{12}$ is given by   \cite{Allen & Romano}
\be\label{SNR constrain}
\text{SNR}_{12}\geq\sqrt2\left(
\text{erfc}^{-1}(2\alpha)-\text{erfc}^{-1}(2\gamma)\right),
\ee
where $\text{erfc}^{-1}(\alpha)$ is the inverse function of
 the complementary error function
$\text{erfc}(z)\equiv\frac2{\sqrt\pi}\int_z^\infty dx\ e^{-x^2}$,
 $\alpha$ is called the false alarm rate,
and $\gamma$ is called the detection rate.
Taking  $\alpha=5\%$ and  $\gamma=95\%$, Eq.~(\ref{SNR constrain})    gives
\be\label{SNR constrain2}
\text{SNR}_{12}\geq3.29.
\ee
Thus, fixing two  parameters out of $(r,\alpha_t,\beta)$,
we can convert (\ref{SNR constrain2}) into
a lower limit on the remaining  parameter of RGW.
Table \ref{constrain correlate} shows the lower limits of
 $\alpha_t$ with the other two  being fixed
for   $T=1$ year.

\begin{table}[htbp]
\caption{The lower limits of $\alpha_t$ with the other two
          being fixed.}
\label{lower limit}
\begin{center}
\begin{tabular}{|p{50pt}|l|l|l|l|l|l|l|l|l|} 
\hline
\setlength{\unitlength}{0.6mm}
\begin{picture}(60,10)
\put(-2.8,10){\line(3,-1){35.5}}
\put(25,4){$\beta$}
\put(3,0){$r$ }
\end{picture}
     &-1.94 & -1.96 & -1.98 & -2 & -2.02 & -2.04 & -2.06 & -2.08 \\
\hline
0.1    & -0.00041 & 0.00190  & 0.00421 & 0.00653 & 0.00884 & 0.01115 & 0.01346 & 0.01577   \\   \hline
0.05    & 0.00074 & 0.00306 & 0.00537 & 0.00768 & 0.00999 & 0.01230 & 0.01461 & 0.01692  \\   \hline
\end{tabular}
\label{constrain correlate}
\end{center}
\end{table}

\section{ Estimation  by integrated signals from a pair}

\subsection{ The integrated output signals }

We now try to
determine the RGW spectrum by  a pair.
Let the sample vector   of the cross-correlated signals
\be \label{Ctime}
\bm C=[C_1,C_2,...,C_N] ,
\ee
where  each  $C_i$ is the cross-correlated, integrated output signal
of  (\ref {correlation signal}),
\be\label{Ci}
C_i  =\int_{-\infty}^\infty df\int_{-\infty}^\infty df'
  \delta_{T_i}(f-f') \tilde s_1^{\, *}(f)   \tilde s_2(f')\tilde Q(f'),
~~~~ i= 1,2,\cdots, N ,
\ee
and $N$ is the number of segments,
and is sufficiently large.
When $T\gg$ the light travel time $L/c\sim 2$\,seconds  between the two detectors of LISA,
non-overlapping  $C_i$ and $C_j$  for $j\neq i$
are statistically independent \cite{Allen & Romano}.
For  each  $i$,
 $C_i$ has the mean $\mu_i=\l\langle C_i\r\rangle$  given by  (\ref {meanmu}),
and the variance $\sigma^2 _i =  \langle C_i^2 \rangle -\mu_i^2 $
given by (\ref {mutosigma}).
In general, $\mu_i$   varies for different $i$, so does   $\sigma^2 _i$.
Denote the mean of $\bm C$  by
$\bm \mu=[\mu_1,\mu_2,...,\mu_N]$,
and the  covariance matrix  by $\bm \Sigma=\l (   \Sigma_{ij}\r)$ with
\be \label{Sigmaij}
\Sigma_{ij}=\l\langle(C_i-\mu_i)(C_j-\mu_j)\r\rangle
   \, , ~~~~ i,j =1,2,\cdots,N ,
\ee
which is diagonal, $\Sigma_{ij}=\delta_{ij}\sigma_j^2$,  by independence.
(Here  $\bm \Sigma$  for a pair
should not be confused with that in Section 4 for a single.)
Explicitly,
\be \label{mui2}
\mu_i = \langle C_i  \rangle
 =   \frac{3 T_i}{10}    m    ,
\ee
\be \label{sigmaij2}
\Sigma_{ij}
  = \delta_{ij}\, b \mu_j,
\ee
where $b \equiv  5/3$ and
\be \label{mdef}
m \equiv   \int_{0}^\infty df  \frac{S^2_h(f)\, \gamma^2(f)}{M(f)}
\ee
is a functional of $S_h(f)$.
We assume that the  PDF of  $\bf C $
is a multivariate Gaussian
\be\label{pdfmult}
f({\bf C}  )=
\frac{1}{(2\pi)^{\frac{N}{2}}\text{det}^{\frac12}[\bm \Sigma ] }
\exp\l\{-\frac12 { (\bm C-\bm\mu) }\ \bm\Sigma^{-1}
 \   \l(\bm C-\bm\mu   \r) ^T\r\} \, ,
\ee
which, by  (\ref{sigmaij2}),  is
 \be \label{pdfmult2}
f ({\bf C}) =
\frac{1}{(2\pi  )^{\frac{N}{2}}\, ( \Pi_i^N b\mu_i)^{\frac12 } }
\exp\l\{ -\frac{1}{2b}  \sum_i^N { \frac{ \l( C_i - \mu_i  \r)^2 }{\mu_i} }
    \r\} \, .
\ee
The likelihood function is,
after dropping an irrelevant  constant $\frac12 N \ln 2\pi$,
 \be\label{like}
\mathcal L   \equiv    -\ln f
= \frac12 \sum_i^N \ln ( b\mu_i)
   + \frac{1}{2b }  \sum_i^N { \frac{ \l( C_i - \mu_i  \r)^2 }{\mu_i} },
\ee
which is  a  functional of the spectrum $S_h$ through  $\mu_i$.
Once the PDF  is chosen,
 an estimator  of the spectrum  is a specification
to give the value $S_h$ for the given data set $\bf C$.
For this,  we shall adopt the ML method.
In general,
$\mathcal L$ can be expanded in a neighborhood of  some spectrum $\bar S_h(f)$,
\bl
 \mathcal L = & \bar{ \mathcal L}
 +
 \sum_{k=1}^N\frac{\partial  \mathcal L}{\partial  S_{h}(f_k)}(S_{h}(f_k)-\bar S_{h}(f_k))
    \nn \\
&   +\frac12
\sum_{k,\,l=1}^N \frac{\partial  ^2 \mathcal L}{ \partial  S_{h}(f_k) \partial S_h(f_l) }
  \l( S_{h}(f_k) - \bar S_{h}(f_k) \r) \l(S_h(f_l)- \bar S_h(f_l)\r).
\el
We  look for  the  most  likely  spectrum $\bar S_h(f)$
at which  $  \mathcal L  $  is minimized
\be\label{MLeq}
\l.\frac{\delta  \mathcal L}{\delta  S_{h} }\r|_{\bar S_h}  =0.
\ee
The first order derivative is  (see  Appendix A for details)
\bl \label{1st0}
\frac{\delta  \mathcal L}{\delta  S_{h} }
= & \frac12  \sum^N_{l}
 \bigg[\frac{1}{ \mu_l} -\frac{C^2_i}{b \mu_i^2} +\frac{1}{b}  \bigg]
   \frac{\delta  \mu_l}{\delta  S_{h} } \nn \\
= & \frac12  \frac{ S_h(f)\, \gamma^2(f)}{M(f)}   \l( 1- \frac{N(f) }{M(f)} \r)
 \l( \frac{N}{m} - \frac{2}{m^2} \sum^N_{i} \frac{C^2_i }{T_i}
                +  \frac{1}{2b^2} \sum^N_{i}  T_i \r) ,
\el
where  $\,m$  is given by (\ref{mdef})
and $N(f)$ is given by (\ref{N}).
The analytical expression of the solution for (\ref{MLeq})
is not available,
and one needs to use numerical methods.
The Newton-Raphson method \cite{Oh1999,Hinshaw2003Apjs,Press1992}
is generally used to  find the ML-estimate of the spectrum.
In many applications,
the  Newton-Raphson method
 is known to converge quadratically in the neighborhood of the root.
For instance, in the  spectral estimation of CMB anisotropies \cite{Oh1999,Hinshaw2003Apjs},
typically 3-4 iterations will be sufficient.
Let $S^{(0)}_h(f)$ be a trial power spectrum,
which can be tentatively chosen as the analytical spectrum (\ref {hspectrum})
with some values of parameters.
In the neighborhood of $S^{(0)}_h(f)$,
the first order derivative of the
likelihood   is expanded  as the following
\be \label{trial}
\frac{\delta  \mathcal L}
{\delta   S_{h}(f)}\bigg|_{S_h(f)}
\simeq
\frac{\delta  \mathcal L}
{\delta   S_{h}(f)}\bigg|_{S^{(0)}_h(f)}
+ \int d f'  \frac{\delta    ^2
\mathcal L}{\delta    S_{h}(f) \delta  S_h(f') } \bigg|_{S^{(0)}_h(f)}
\l(S_h(f')-S^{(0)}_h(f')\r) =0.
\ee
As an approximation,
$\frac{\delta^2 \mathcal L}{ \delta   S_{h}  \delta S_h }$
is replaced by its expected value, i.e,  the Fisher matrix,
\bl \label{FfactorF}
\mathcal F(f, f')
& = \l[ \frac{ S_h(f')\, \gamma^2(f')}{M(f^{\,'})}
  \l( 1-  \frac{N(f') }{M(f')}  \r)\r]
  \l[\frac{ S_h(f)\, \gamma^2(f)}{M(f)}
  \l( 1-  \frac{N(f) }{M(f)}  \r)\r] \frac{1}{2}
  \l(  \frac{N}{m^2} +  \frac{ 9}{25 m } \sum_l^N T_l \r),
\el
(see  Appendix A for  the derivation).
However, this Fisher matrix is degenerate and has no inverse,
and one will not be able to invert Eq.(\ref{trial})
to get an estimated spectrum.
This is because
the signal $C_i$ constructed  in (\ref{Ci}) is an integration over frequency,
as is $\mu_i$.
On the other hand, for spectrum estimation,
one needs to assign a value  $S_h(f_j)$ at each frequency   $f_j$.
Thus, we conclude that
$C_i$ will not help to estimate the RGW spectrum by a pair,
even though it is useful for detection of an RGW signal.

\subsection{Parameter estimation  in a Bayesian approach  }

We shall be able to use  $C_i$ to estimate one parameter of RGW
in a Bayesian approach.
Consider   the PDF  as in Eq.~(\ref{pdfmult2}),
\be\label{PDFcross}
f ({\bf C};\bm\theta) =
\frac{1}{(2\pi  )^{\frac{N}{2}}\, ( \Pi_i^N b\mu_i (\bm\theta))^{\frac12 } }
\exp\l\{ -\frac{1}{2 b}  \sum_i^N {\frac{ \l( C_i -\mu_i (\bm\theta)\r)^2}{\mu_i(\bm\theta)} }
    \r\} \, ,
\ee
where  $\bm \mu(\theta)$ and  $\bm \Sigma(\bm\theta)$
as in (\ref{mui2}) and (\ref{sigmaij2}) respectively
now depend on the RGW parameters through the theoretical spectrum $S_h$,
and $\bm \theta$ denotes the RGW parameters which are  random variables
since they are some functions of the data set   \cite{Kay}.
We  adopt  the  unbiased estimation,  which assumes that
  the average value of an estimator  of the   parameters $\bm  \theta$
is regarded as  its true value.
Using the  ML  method,
the likelihood function $ \mathcal L= - \ln f( \bm\theta) $ can also be
Taylor expanded  around certain values $  \bm {\bar \theta}$
\[
\mathcal L = \bar{ \mathcal L}
+\sum_a  \frac{\partial\mathcal L}{\partial\theta_a}  \bigg|_{\bm {\bar \theta}} \,
           (\theta_a -\bar \theta_a)
+\frac12 \sum_{a,b} \,
\frac{\partial^2\mathcal L}{\partial \theta_a  \partial \theta_b }  \bigg|_{\bm {\bar \theta}}
(\theta_a -\bar \theta_a)(\theta_b -\bar \theta_b)+\,  \cdots\,.
\]
Now we require   $  \bm {\bar \theta}$ to be
the ML estimator,
at which
\be
\frac{\partial \mathcal L }{\partial\theta_a} \,  \bigg|_{ \bm {\bar \theta} }
    =0,  ~~~~a=1,2,3.
\ee
As is known \cite{Kay},
when $N$ is large enough,
the second order derivative at $\bm {\bar  \theta} $
is equal to its average value,
\[
\mathcal F_{ab} \equiv
 \bigg \langle \frac{\partial^2\mathcal L}{\partial \theta_a  \partial \theta_b }\,
 \bigg|_{\bm {\bar \theta} }              \bigg \rangle
= \frac{\partial^2\mathcal L}{\partial \theta_a  \partial \theta_b }
    \bigg|_{\bm {\bar \theta} } ,
    ~~~~a,b=1,2,3,
\]
so that   in the neighborhood   of $\bm {\bar  \theta }$,
  the PDF of (\ref{PDFcross}) becomes
the following Bayesian PDF
in the parameter space
\be\label{MLestimate}
f ( \bm \theta)  \propto    \exp{[-\mathcal L] }
 \propto
\exp\l[-\frac12( \bm\theta-\bm{ \bar \theta})
{\bf \mathcal F} (\bm\theta -\bm {\bar \theta })^T\r] \, ,
\ee
which  is   approximately Gaussian
 in a neighborhood of $\bm {\bar  \theta} $.
For  detailed derivation see appendix 7B of Ref.\cite{Kay}.

The likelihood function follows (\ref{PDFcross}) as
\be
\mathcal L(\bm C;\bm\theta)\equiv    -\ln f ({\bf C};\bm\theta)
= \frac12 \sum_i^N \ln \l( b\mu_i (\bm\theta) \r)
   + \frac{1}{2 b }\sum_i^N { \frac{ \l( C_i - \mu_i (\bm\theta)\r)^2 }{\mu_i (\bm\theta)} }.
\ee
To estimate $\bm   \theta$,
one needs  the first order derivative     (See  Appendix A),
\bl \label{parameq}
\frac{\partial   \mathcal L}{\partial   \theta_a}
= & \int df \frac{ S_h(f)\, \gamma^2(f)}{M(f)}   \l( 1- \frac{N(f) }{M(f)} \r)
 \frac{\partial   S_{h}(f)}{\partial  \theta_a}
 \l( \frac{N}{m}   - \frac{2}{m^2} \sum^N_{i} \frac{C^2_i }{T_i}
                +  \frac{1}{2  b^2} \sum^N_{i}  T_i \r)  ,
\el
and the $3\times 3$ Fisher matrix
\bl \label{FFish}
\mathcal F_{ab}
=  & \l( \int_{0}^\infty df '
                \frac{ S_h(f')\, \gamma^2(f')}{M(f')}
                \l( 1- \frac{N(f') }{M(f')} \r)\,
                   \frac{\partial   S_{h}(f')}{\partial   \theta_a} \r) \nn \\
&   \times  \l(   \int_{0}^\infty df
           \frac{ S_h(f)\, \gamma^2(f)}{M(f)}
                \l( 1- \frac{N(f) }{M(f)} \r)\,
                \frac{\partial   S_{h}(f)}{\partial   \theta_b} \r)
      2    \l(  \frac{N}{m^2} + \frac{9}{ 25 m} \sum_i^N T_i \r)\, .
\el
However, this  Fisher matrix is  degenerate
and has no inverse.
Thus,
one can not determine the whole set $(r,\beta,\alpha_t)$ simultaneously.
What one can do is to estimate only one of the RGW parameters,
while the other two parameters   are fixed at certain values,
or marginalized.
Note that this method can not determine the correlation between
two parameters, which will be given by another method in Section 8.3 later.
For the former case, one gets the conditional PDF,
and for the latter,
one integrates the PDF of (\ref{PDFcross}) over  $\theta_b$ and  $\theta_c$,
and gets the  marginal PDF for  $\theta_a$
\be
f ({\bf C};\theta_a)  \equiv \int  \int f({\bf C};\bm\theta)   d \theta_b  d\theta_c
\ee
and the marginal likelihood function
$\mathcal L(\bm C; \theta_a)\equiv  -\ln f ({\bf C}; \theta_a)$.
With these  specifications,
one can estimate the parameter $\theta_a$.
Let $ \theta ^{(0)}_a $ be a trial  parameter.
We expand the first order derivative of $\mathcal L$,
 conditional or marginalizing,
around $\theta ^{(0)}_a$
\be \label{trialtheta}
\frac{\partial\mathcal L}{\partial \theta_a}\bigg|_{\bm \theta  }
\simeq
\frac{\partial\mathcal L}{\partial \theta_a}\bigg|_{\bm \theta ^{(0)}  }
+     \frac{\partial^2 \mathcal L}{ \partial   \theta_a \partial \theta_a }
  \bigg|_{ \bm \theta ^{(0)} }
   \l(\theta_{a} -\theta_a^{(0)}  \r) =0.
\ee
Replacing  $\frac{\partial^2  \mathcal L}  {\partial\theta_ a \partial\theta_ a} $
by the $(aa)$ element of Fisher matrix
$\mathcal F_{aa}  \equiv \l \langle \frac{\partial^2}
  {\partial\theta_ a \partial\theta_ a}  \mathcal L \r \rangle $,
one gets
\be\label{esttheta}
\theta_{a} = \theta_a^{(0)}
- \mathcal F_{aa}^{-1}\frac{\partial\mathcal L}{\partial \theta_a}
   \,    \bigg|_{\bm \theta ^{(0)} }  .
\ee
By iteration, one will obtain the estimate of   $\theta_{a}$.
This is a general formula to estimate one parameter.
The Fisher matrix $\mathcal  F_{ab}$  also
gives the standard error of the estimated parameter.
When the data sample is sufficiently large,
one can take the equality in the Cramer-Rao lower bound \cite{Kay}
\be\label{varianceCRL}
\sigma^2_{\theta_a}     = \mathcal  F^{-1}_{aa}(f)
\ee
where $\mathcal  F^{-1}_{aa} $ is
evaluated at the ML-estimate parameter $\theta_a$
 that has been obtained.

In fact,
when noise is dominant over the RGW signal,
the estimate of    $r$  can be also obtained  analytically.
By the property $\,m  \propto r^2 $ implied by  (\ref{mdef}),
one can write
$\,m (r)=r^2 m(r=1)$ with  $\beta$ and $\alpha_t$ being  fixed in  $\,m(r=1)$.
Setting (\ref{1st0}) to zero, and solving for $\,m$, one has the positive root
\bl\label{mC}
\bar m
= & \frac{25}{9}\frac{1}{\frac{1}{N} \sum_i^N T_i }
  \l( -1 + \sqrt{ 1+ \frac{36}{25} \frac{1}{N^2} \l( \sum_j^N T_j \r)
      \l( \sum_i^N \frac{C_i^2}{T_i} \r) }  \r) \, ,
\el
 from which one obtains   the analytical ML-estimate of $r$ as the following
\be\label{r-estimate}
r= \frac{1}{ \sqrt{m(r=1)} }\sqrt{\frac{ 25/9 }{\frac{1}{N} \sum_i^N T_i }
  \l( -1 + \sqrt{ 1+\frac{36}{25N^2}  \l(  \sum_j^N T_j  \r)
     \sum_i^N  \frac{C_i^2}{T_i}}  \r)}   \, .
\ee
However, no analytical   ML-estimates  are available for  $\beta$ and $\alpha_t$.
Still one can estimate   $\beta$ and $\alpha_t$
in a  manner simpler than  (\ref{esttheta}).
Let $\theta_a$ be    $\beta$ and
 \be\label{thetanewttheta}
\mathcal M [\theta_a]
    \equiv  \int_{0}^\infty df  \frac{S^2_h(f)\, \gamma^2(f)}{M(f)}- \bar m  =0,
\ee
in which  $r$ and $\alpha_t$ are fixed.
We use the  Newton-Raphson method  by iterations
as before.
Write $\mathcal M [\theta_a] $ as
\be\label{mthetas}
 \mathcal M [\theta_a] \simeq \mathcal M [\theta_a^{(0)} ]
   +  \frac{\partial  \mathcal M}{\partial  \theta_a } \bigg|_{\theta_a ^{(0)}} \,
   \l( \theta_a  - \theta^{(0)}_a \r) =0,
\ee
where   $\theta^{(0)}_a$ is a trial value and
\be
\frac{\partial  \mathcal M}{\partial  \theta_a }
        =  \int_{0}^\infty df
                \frac{2 S_h(f)\, \gamma^2(f)}{M(f)}
                \l( 1- \frac{N(f) }{M(f)} \r)\,
                   \frac{\partial   S_{h}(f)}{\partial   \theta_a}.
\ee
One solves (\ref{mthetas}) and gets the estimate
\be \label{thetaest}
 \theta_a  =  \theta_a^{(0)}
  -  \frac{\partial  \mathcal M}{\partial  \theta_a } \bigg|_{\theta_a^{(0)}}^{-1} \,
  \mathcal M [\theta_a^{(0)} ] .
\ee
Similarly, the estimation of  $\alpha_t$ can be also done.
Note that since the filter function $\hat Q$ in (\ref{Qf})
contains the theoretical spectrum $S_h(f)$,
this ML-estimation method is essentially a technique of matched filter
       \cite{Gair2013}.

We  perform a  numerical simulation  to estimate  $r$,
  using (\ref{esttheta}).
For    $r=0.1$, $\alpha_t=0.016$ and $\beta=-2.016$ with SNR$_{12}$=179,
we use the PDF (\ref{pdfmult2}) to generate cross-correlated data stream
numerically.
We take  $T \simeq 3$\,hours $\sim$$10^4$\,s for one segment,
with total observation duration   $\sim 1$ year,
and  number of segments  $n \sim 3\times10^3$.
Then we   estimate  $r$ numerically by (\ref{esttheta}),
and after five steps of iteration,
$r$ converges to $r_{ML}=0.1011$.

According to (\ref{MLestimate}) and (\ref{varianceCRL}),
the standard  deviation   is
$\sigma_{\theta_a}\equiv 1/ \sqrt {\mathcal F_{aa}}$.
If the estimation is required to be  at the $95\%$   confidence level (cl),
$0.95 = \frac{2}{\sqrt \pi}\int_0^{\Delta \theta_a  /(\sqrt{2}\sigma_{\theta_a}) }
e^{- t^2}dt$,
the resolution of the estimated parameter $\theta_a$
will  be
\be
\Delta\theta_a
= 1.96\sigma_{\theta_a} \ \  {\rm\ at\ } 95\% \,\,{\rm cl} \, .
\ee
Table \ref{resolution abr} lists
 the resolution of $r$, $\alpha_t$ and $\beta$,  separately,
 and
the corresponding values of SNR$_{12}$ ($\geq 3.29$).
\begin{table}[htbp]
\Large
\caption{ Resolution of $r$, $\alpha_t$ and  $\beta$ separately
  at $95\%$ cl for a pair}
\begin{center}
\label{resolution abr}
\begin{tabular}{|c |c|c|| c| c|c||c|}
\hline
$r$&$\alpha_t$ & $\beta$ & $\Delta r/r$&$\Delta\alpha_t$ & $\Delta\beta$&  SNR$_{12}$
\\
\hline
0.05&0.01&-2.016&
$2.58\times10^{-2}$  & $4.29 \times10^{-5}$ &$3.72\times10^{-4}$  & 4.36
\\
\hline
0.05&0.015&-2.016&
$1.46\times10^{-2}$  & $2.42 \times10^{-5}$ &$2.10\times10^{-4 }$  & 72.8
\\
\hline
0.1&0   & -1.93 &
$2.56\times10^{-2}$& $4.28 \times10^{-5}$ &$3.70\times10^{-4}$  &  8.34
\\
\hline
0.1&0.01&-2.016&
$2.55\times10^{-2}$  & $4.26 \times10^{-5}$ &$3.69\times10^{-4}$  & 8.62
\\
\hline
0.1&0.01& -1.93&
$1.46\times10^{-2}$  & $2.40 \times10^{-5}$ &$2.09\times10^{-4}$  &  390
\\
\hline
0.1&0.016&-2.016&
$9.72\times10^{-3}$  & $1.61\times10^{-5}$ &$1.40\times10^{-4}$  & 179
\\
\hline
\end{tabular}
\end{center}
\end{table}

\section{Spectral estimation by ensemble average of a pair}

To estimate the RGW spectrum,
we turn to the method of ensemble  averaging of data from a pair.
Consider the output signals $\tilde s_1(f)$,  $\tilde s_2(f)$
in frequency space
from (\ref{output individually1}) and (\ref{output individually2}) respectively.
Since the noises are uncorrelated,
the ensemble average
$\l\langle s_1(f)\tilde s_2(f') \r\rangle$ is
given by (\ref{ensemble average of s1s2}).
In practice,
when there are $n$ independent sets of observational data,
each being $(s_1(f)\tilde s_2(f'))_i$,
they can  form the sample mean,
\be\label{samplemean}
\l\langle s_1(f)\tilde s_2(f') \r\rangle_t
   \equiv \frac{(s_1(f)\tilde s_2(f'))_1+\cdots+(s_1(f)\tilde s_2(f'))_n }{n} \, ,
\ee
which represents the ensemble average
when the independent sets of data  are large enough.
Thus (\ref{ensemble average of s1s2}) becomes
\be\label{s1s2toSpectrum2}
\langle \tilde s_1(f)\tilde s_2(f')\rangle_t
=\frac{1}{2}\delta (f-f')  S_h(f)\mathcal R_{12}(f).
\ee
In practical analysis,
we can replace $\delta(f-f')$ by its discrete  form  in (\ref{deltaij3}),
\be\label{s1s2toSpectrum3}
\langle \tilde s_1(f_i)\tilde s_2(f_j) \rangle_t
= \frac{\delta_{ij}} {2 \Delta f} S_h(f_i)\mathcal R_{12}(f_i) .
\ee
Solving  Eq.~(\ref{s1s2toSpectrum3}),
one obtains an estimate of the RGW spectrum  by a pair
\be\label{s1s2toSpectrum4}
\bar S_h(f_i)
= \frac{ 2\Delta f }{\mathcal R_{12}(f_i)}
\l\langle \tilde s_1(f_i)\tilde s_2(f_i)\r\rangle_t  .
\ee
(\ref{s1s2toSpectrum4}) is the main formula in our paper
to estimate the spectrum from a pair.
As an advantage,
it does not require a priori knowledge of the  noise spectrum,
in contrast to (\ref{ShEstimator}).
In the ensemble averaging method,
$\langle \tilde s_1(f_i)\tilde s_2(f_j) \rangle_t $
as the basic quantity
does not involve integration over frequency.

Using (\ref{s1s2toSpectrum4}),
we   conduct a numerical simulation to examine its feasibility.
First,  we construct  the vector of RGW output response
${\bf \tilde h_o}(f_i)\equiv [\tilde h_1(f_i), \tilde h_2(f_i)]$ with
$i=1,2,\cdots,N$,
where each of $\tilde h_1(f_i)$ and $\tilde h_2(f_i)$
is defined as in Eq.~(\ref{hij}).
The mean and  variance are
given in Eqs.~(\ref{Sh normalization2}) and (\ref{ensemble average of s1s2}),
and the   corresponding PDF is
\be \label{PDFRGW}
f\l({\bf\tilde h_o}(f_i)\r)=
\frac{1}{(2\pi)^{\frac{N}{2}}\text{det}^{\frac12}[\bm{\Sigma}_{(h)}(f_i)]}
\exp\l\{-\frac12 {\bf \tilde h_o}(f_i)
\l[\bm{\Sigma}_{(h)}(f_i)\r]^{-1}
\l[{\bf \tilde h_o}(f_i)\r]^T\r\},\ i=1,2,\cdots,N \ ,
\ee
where the covariance matrix is
\be\label{Covarianceh1h2}
\bm{\Sigma}_{(h)}(f_i)\equiv \frac{1}{2  \Delta f} S_h(f_i)\left(
                              \begin{array}{cc}
                                  \mathcal R(f_i)_1
                                & \mathcal R_{12}(f_i) \\
                                  \mathcal R_{12}(f_i)
                                &  \mathcal R(f_i)_2    \\
                              \end{array}
                            \right)
,\ i=1,2,\cdots,N \ .
\ee
Here $ \mathcal R(f_i)_1$ and $ \mathcal R(f_i)_2$
are transfer function of the interferometer $1$ and $2$ respectively,
and we can assume $R_1\simeq R_2$,
and $\mathcal R_{12}$ is the transfer function
for the pair defined in (\ref{correlated transfer function}).
The  inverse matrix of (\ref{Covarianceh1h2})  is
\be
[\bm{\Sigma}_{(h)}(f_i)]^{-1}= \frac {2 \Delta f}{ S_h(f_i)}
           \frac{1}{\mathcal R^2(f_i)-\mathcal R_{12}^2(f_i)}
           \left(
                           \begin{array}{cc}
                \mathcal R(f_i)
                        &
              - \mathcal R_{12}(f_i)
                                \\
              - \mathcal R_{12}(f_i)
                           &
              \mathcal R(f_i)  \\
                      \end{array}
                            \right)
  \ .
\ee
Similarly, for the noise in the pair,
we write  the  noise  vector
${\bf \tilde n}(f_i)\equiv  [\tilde n_1(f_i), \tilde n_2(f_i)]$.
The mean and covariance are in (\ref{noise emsemble}) and (\ref{moisesp}),
and the PDF is
\be \label{PDFNoise}
f\l({\bf\tilde n}(f_i)\r)=
\frac{1}{(2\pi)^{\frac{N}{2}}\text{det}^{\frac12}[\bm{\Sigma}_{(n)}(f_i)]}
\exp\l\{-\frac12 {\bf \tilde n}(f_i)
\l[\bm{\Sigma}_{(n)}(f_i)\r]^{-1}
\l[{\bf \tilde n}(f_i)\r]^T\r\} \ ,
\ee
where  the covariance matrix is
\be\label{Covariancen1n2}
\bm{\Sigma}_{(n)}(f_i)\equiv \frac{  S_n(f_i)}{2  \Delta f}  \left(
                              \begin{array}{cc}
                                1&0 \\
                               0 & 1 \\
                              \end{array}
                            \right)
  \ ,
\ee
which is diagonal since the  noises in the pair are  uncorrelated.
Based on the above construction,
the joint PDF for the total output signal is
given by (\ref{gaussone})
with the covariance matrix
$\bm{\Sigma}(f_i)=\bm{\Sigma}_{(h)}(f_i)+\bm{\Sigma}_{(n)}(f_i)$.

\begin{figure}[htbp]
\centering
  \includegraphics[width=0.8\columnwidth]{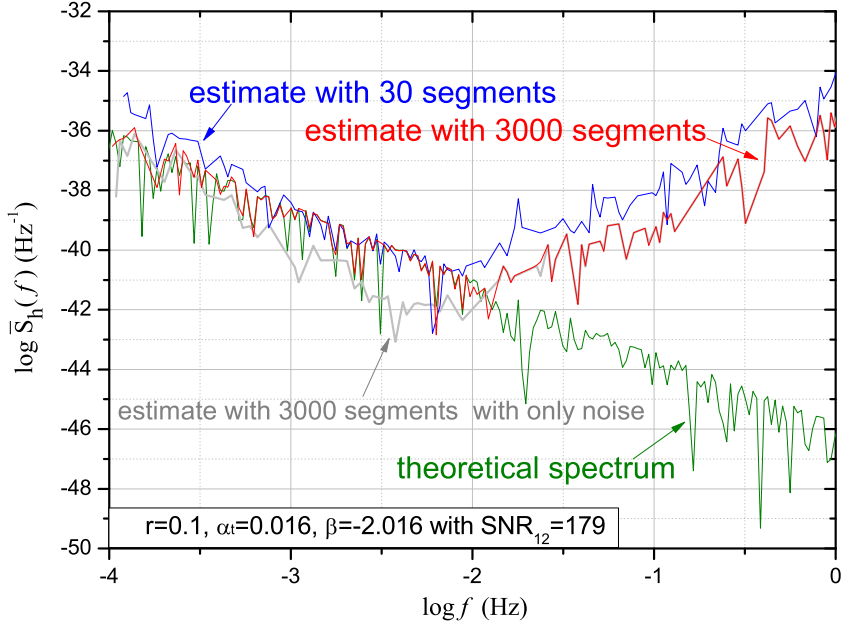}\\
  \caption{
  The estimated spectrum by a pair.
  The cases with only the  theoretical spectrum and
  simulated  noise are also shown.
  }\label{ShCorSu}
\end{figure}

We use PDFs  (\ref{PDFRGW}) and (\ref{PDFNoise}) to numerically
generate the output response and noise of a pair.
We need the  data set  of $n$  segments
$({\bf \tilde s}(f_i))_{ 1},\cdots, ({\bf \tilde s}(f_i))_{ n}$.
One can take  one typical segment of the data stream
with  a period of $T \simeq 3$\,hours $\sim$$10^4$\,s,
with total observation duration    $\sim 1$ year,
and   number of segments   $n \sim 3\times10^3$.
According to the PDF of (\ref{PDFRGW}) and (\ref{PDFNoise})
with the specific  variance $ \bm{\Sigma}$
by taking RGW with   $r=0.1$, $\alpha_t=0.016$, $\beta=-2.016$
with SNR$_{12}=179$,
we  numerically  generate $3\times10^3$
independent sets of random data streams ${\bf \tilde s}(f_i)$ ($i=1,\cdots,N$).
Substituting  these generated   data streams into  Eq.~(\ref{s1s2toSpectrum4}),
we obtain  the estimated RGW spectrum $\bar S_h(f_i)$
for $n=30$ and $n= 3000$, as shown in Fig.~\ref{ShCorSu}.
For illustration,
the theoretical spectrum and the simulated  noise are also shown.
The estimation depends on the length  of simulated data.
A longer length $n$ of data gives a better estimate.
In an ideal case of infinitely long data  length,
the off-diagonal elements of the covariance matrix for
practical noise signal would be 0,
and the RGW signal at all relevant frequencies  could  be detected.
But with a finite size of data,
the estimation will be limited when the  noise is large.
As Fig.~\ref{ShCorSu} shows,
the estimate    (\ref{s1s2toSpectrum4}) at high frequencies
is actually contributed by the nonzero off-diagonal elements of noise, i.e,
$\bar S_h(f)\sim$$\frac{ 2\Delta f \langle\tilde n_1(f_i)\tilde n_2(f_i)
\rangle}{\mathcal R_{12}(f_i)}$ at $f>10^{-2}$\,Hz,
because of a finite length of data.

\section{Estimations by correlation of un-integrated signals from a pair }

In this section,
we adopt a method of correlation of un-integrated signals
to estimate the spectrum and parameters of RGW as suggested by Ref.~\cite{Seto2006}.
By dividing the whole frequency range  of the data
into many small segments,
the mean value of the correlation variable over each segment
is taken as the representative point for the segment.
As an approximation, the method is able to give the estimate of
the RGW spectrum,
as well as the three parameters of RGW,
improving that  in Section 6.
We assume that the output data are sufficient for this purpose.

\subsection{A correlation variable of un-integrated signals}

In analogy with (3.4) in Ref.~\cite{Seto2006},
we divide a positive frequency range into $N$ segments,
and the $i$-th segment $F_i$ $(i=1,2,...,N)$ of width $\delta f_i$
has a center frequency $f_i$.
For instance, the frequency range is taken as $(10^{-4} \sim 1)$ Hz for LISA,
  $N \sim  10^4$, and $\delta f\sim   10^{-4}$\,Hz.
A correlation variable is defined in each segment $F_i$ as
\be\label{corZi}
 Z_i\equiv  \sum_{f\in F_i}
 \Delta f \,
 \tilde s^*_1(f)\tilde s_2(f) , \,\,\, i=1,2,...,N
\ee
where the frequency resolution $\Delta f =1/T_1 \ll \delta f_i$
with $T_1$ being the observation period, say $\Delta f \sim 10^{-6}$\,Hz,
so that each segment contains a large number of Fourier modes.
(Notice that (3.4) in Ref.\cite{Seto2006} should have a factor of  $\Delta f$
for consistency of dimension.)
The mean of $Z_i$ is
\be\label{mui}
\mu_i=  \langle Z_i\rangle=  \sum_{f\in F_i}\Delta f
    \frac{1}{2}\delta(f-f) S_h(f)\mathcal R_{12}(f),
\ee
where   (\ref{noise uncorrelated}) and (\ref{ensemble average of s1s2})  are used.
Using the formula (\ref{deltaij3}) to replace the Dirac delta function
by its discrete form,
(\ref{mui})  can be written as the following
\be
\mu_i= \sum_{f\in F_i}\Delta f \frac{1} {2 \Delta f} S_h(f)\mathcal R_{12}(f) .
\ee
The summation  can be approximately replaced as
\be\label{MeanCorNew}
\mu_i
= \frac{\delta f_i} {2\Delta f} S_h(f_i)\mathcal R_{12}(f_i)  ,
\,\,\, i=1,2,...,N
\ee
where $S_h(f_i)\mathcal R_{12}(f_i)$ is the mean value over the $i$-th segment,
as suggested in  Ref.~\cite{Seto2006}.
To keep the error small in this approximation,
$\delta f_i$ should  be sufficiently  small
so that the mean value of the function
represents the summation function within $\delta f_i$.
(We note that the overlapping reduction function
$\gamma_{12}(f)$ defined in Ref.~\cite{Seto2006}
is related to our  $\mathcal R_{12}(f) $ by
$\gamma_{12}(f)= \frac{5}{2}\mathcal R_{12}(f)$,
which together with (\ref{hspectrum}) leads to
$\mu_i = \frac{\delta f_i} {\Delta f}
\frac{3H_0^2}{20\pi^2}\frac{\Omega_g(f)}{f^3}\gamma_{12}(f) $,
 the same as (3.5) in Ref.~\cite{Seto2006}.)

The variance of $Z_i$ is
\be
\sigma_i^2
=\l\langle \Big(Z_i-\langle Z_i\rangle\Big)^2\r\rangle
=\langle Z_i^2\rangle
-\langle Z_i\rangle^2 .
\ee
By noting that $\int_{-\infty}^{+\infty}df'\delta(f-f')$  is equivalent to
 $2\sum_{f'\in F_j}\Delta f\delta(f-f')$,
(\ref{corZi}) can be written as
\be
 Z_i=    \sum_{f\in F_i}\Delta f
 \l[2\sum_{f'\in F_j}\Delta f\delta(f-f')\r]
 \tilde s^*_1(f)\tilde s_2(f') ,
\ee
and  the  variance is written as
\bl
\sigma_i^2
& =4\l\langle
\sum_{f\in F_i}\Delta f\sum_{f'\in F_j}\Delta f\delta(f-f')
 \tilde s^*_1(f)\tilde s_2(f')
\sum_{k\in F_i}\Delta f\sum_{k'\in F_j}\Delta f\delta(k-k')
 \tilde s^*_1(k)\tilde s_2(k')   \r\rangle
     - \mu_i^2 \, .
\el
By similar calculations leading to (\ref{variance correlation1}),
we obtain the following result
\be\label{VarCorNew}
\sigma_i^2
= \frac{1}{8}\frac{\delta f_i} {\Delta f}   M(f_i) , \,\,\, i=1,2,...,N
\ee
where $M(f)$ is defined in (\ref{MS12}).
For large noise, $ M (f) \simeq  S_{1n} (f) S_{2n} (f)$,
\be\label{VarNoiseCorNew}
\sigma_i^2
\simeq
\frac{1}{8}\frac{\delta f_i} {\Delta f}
S_{1n}(f_i)S_{2n}(f_i) ,
\ee
which is the same as (3.6) in Ref.~\cite{Seto2006}.

SNR of each segment for this correlation is defined as \cite{Seto2006}
\be
\text{SNR}^2_{i}\equiv \frac{\mu_i^2}{\sigma_i^2}
=2\frac{\delta f_i} {\Delta f}\frac{\mathcal R_{12}^2(f_i) S_h^2(f_i)}{ M(f_i)} ,
\ee
and summing up all segments yields the total SNR
\be
\text{SNR}^2_{\text{C}}
=2\sum_{i=1}^N\frac{\delta f_i} {\Delta f}\frac{\mathcal R_{12}^2(f_i) S_h^2(f_i)}{ M(f_i)} .
\ee
Replacing $\Delta f$ with $1/T_1$,
the summation $\sum_{i=1}^N\delta f$ with integration $\int_0^\infty df$,
one has the SNR over an observation period $T_1$ as
\be\label{SNRC}
\text{SNR}_{\text{C}}
=\sqrt{2 T_1 }
\l[\int_0^\infty df\frac{\mathcal R_{12}^2(f) S_h^2(f)}{ M(f)}\r]^{1/2}.
\ee
When the whole observation duration $T$ consists of many observation periods,
we can use $T$ to replace $T_1$ in the above formula.
This result is consistent with (\ref{SNR max correlation})
by noting that $\mathcal R_{12}(f)=\frac{3} {10}\gamma(f)$.

\subsection{Spectrum estimation}

\begin{figure}[htbp]
\centering
  \includegraphics[width=0.8\columnwidth]{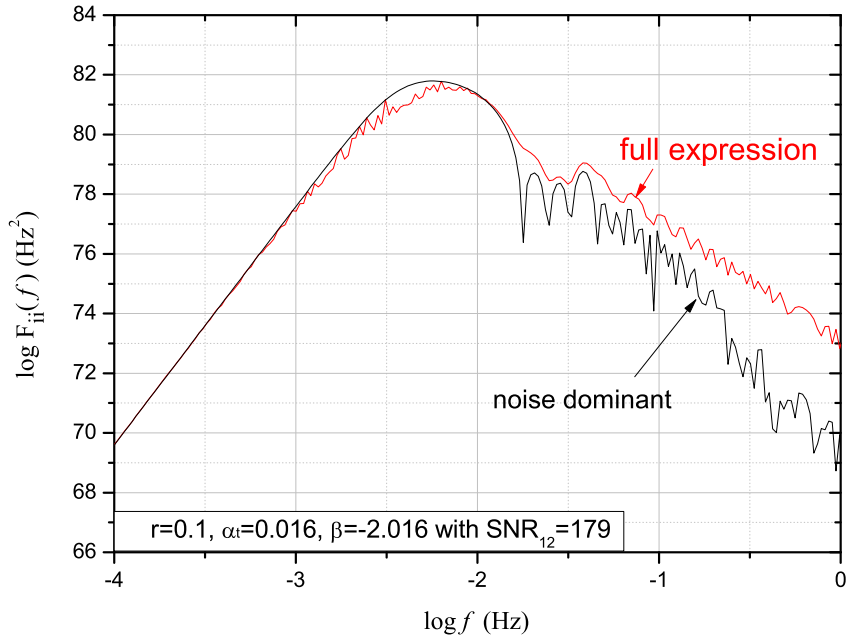}\\
  \caption{
The diagonal element $  F_{ii}$ to estimate spectrum
for both the full expression (\ref{FisherCorNew}) and
the noise dominant (\ref{FisherCorNewNoise}) cases.
   }\label{FisherDiaSeto}
\end{figure}

Since  $\delta f_i/\Delta f\gg1$, according to the central limit theorem,
$Z_i$ can be   described by a Gaussian distribution,
and the PDF   for ${\bf Z}\equiv   [Z_1, Z_2,...,Z_N ]$ is
\be\label{PDFcorNew}
f ({\bf Z}) =
\frac{1}{(2\pi  )^{\frac{N}{2}}\, ( \Pi_i^N \sigma_i^2)^{\frac12 } }
\exp\l\{ -\frac{1}{2}  \sum_i^N { \frac{ \l( Z_i - \mu_i  \r)^2 }{\sigma_i^2} }
    \r\} \, .
\ee
The likelihood functional is,
after dropping an irrelevant  constant $\frac12 N \ln 2\pi$,
 \be\label{CorNewlike}
\mathcal L   \equiv    -\ln f  ({\bf Z})
= \frac12 \sum_i^N \ln ( \sigma_i^2)
   + \frac{1}{2 }  \sum_i^N { \frac{ \l( Z_i - \mu_i  \r)^2 }{\sigma_i^2} },
\ee
which is  a  functional of the spectrum $S_h$ through  $\mu_i$.
We  look for   the  most  likely  power  spectrum $\bar S_h$
at which
 $\frac{\partial\mathcal L}{\partial  S_{h} }\big|_{\bar S_h}  =0$.
The first order derivative is
\be
\frac{\partial  \mathcal L}{\partial  S_{h}(f_j) }
= \frac12 \sum_i^N \bigg[
\frac{1}{ \sigma_i^2}\frac{\partial\sigma_i^2}{\partial  S_{h}(f_j) }
-\frac{ \l( Z_i - \mu_i  \r)^2 }{(\sigma_i^2)^2} \frac{\partial\sigma_i^2}{\partial  S_{h}(f_j) }
-\frac{2 \l( Z_i - \mu_i  \r) }{\sigma_i^2}\frac{\partial\mu_i}{\partial  S_{h}(f_j) }
    \bigg] .
\ee
Plugging (\ref{MeanCorNew}) and (\ref{VarCorNew}) into the above,
by  $\frac{\partial S_h(f_i)}{\partial S_h(f_j)}=\delta_{ij}$,
one has
\be\label{LikePShCorNew}
\frac{\partial  \mathcal L}{\partial  S_{h}(f_i) }
=
 \frac{N(f_i)}{S_{h}(f_i)M(f_i)}
-\frac{ 8\l( Z_i - (\frac{\delta f_i} {2\Delta f} S_h(f_i)\mathcal R_{12}(f_i)  )  \r)^2 }
{\frac{\delta f_i} {\Delta f} S_{h}(f_i)\l[M(f_i)\r]^2} N(f_i)
-\frac{4 \mathcal R_{12}(f_i)
\l( Z_i - (\frac{\delta f_i} {2\Delta f} S_h(f_i)\mathcal R_{12}(f_i)  ) \r) }{M(f_i)},
\ee
where  $N(f)$ is defined in (\ref{N}).
In the neighborhood of a trial spectrum $S^{(0)}_h(f)$,
 the first order derivative   is expanded  as the following
\be \label{trialCorNew}
\frac{\partial  \mathcal L}
{\partial   S_{h}(f_i)}\bigg|_{S_h(f_i)}
\simeq
\frac{\partial  \mathcal L}
{\partial   S_{h}(f_i)}\bigg|_{S^{(0)}_h(f_i)}
+ \sum_{k=1}^N \frac{\partial^2
\mathcal L}{\partial    S_{h}(f_i) \partial  S_h(f_k) } \bigg|_{S^{(0)}_h(f)}
\l(S_h(f_k)-S^{(0)}_h(f_k)\r) =0.
\ee
As an approximation,
$\frac{\partial^2 \mathcal L}{ \partial   S_{h}  \partial S_h }$
is replaced by its expected value, i.e,
 the Fisher matrix which  by a formula  similar to (\ref{Likelyhood2})
 is given by
\bl\label{FisherShij}
F_{ij} =
 \l\langle\frac{\partial ^2\mathcal L} {\partial  S_{h}(f_i)\partial S_h( f_j)}\r\rangle
=&
\sum_{k,l=1}^N\frac{\partial  \mu_k}{\partial   S_{h}(f_i)}
\frac{\delta_{kl}}{\sigma_k^2}\frac{\partial   \mu_l}{\partial  S_{h}( f_j)}
+\frac12\sum_{k,l,m,r=1}^N\l(\frac{\delta_{kl}}{\sigma_k^2}
\delta_{lm}\frac{\partial   (\sigma_l^2)}{\partial  S_{h}( f_i)}
\frac{\delta_{mr}}{\sigma_m^2}
\delta_{rk}\frac{\partial  (\sigma_r^2)}{\partial  S_{h}(f_j)}\r).
\el
Substituting   (\ref{MeanCorNew}) and (\ref{VarCorNew}) into the above
yields
\bl\label{FisherCorNew}
F_{ij}
=&
\l[
\frac{\delta f_i} {\Delta f}
\frac{2\mathcal R_{12}^2(f_i)  }{M(f_i)}
+\frac{2 N^2( f_i)}{S^2_h(f_i)M^2(f_i)}
\r]
\delta_{ij}.
\el
It is remarked that
the Fisher matrix is not degenerate,
in contrast to (\ref{FfactorF}) which is degenerate.
In  the approximation of large noise,
one has $M(f)\simeq S_{1n}(f)S_{2n}(f)$, $N(f)=\frac12 S_h(f) \frac{\partial{M(f)}} {\partial S_h(f)}\simeq0$, and
(\ref{FisherCorNew})  reduces to that used in Ref.\cite{Seto2006}
\bl\label{FisherCorNewNoise}
F_{ij}
\simeq  &
\l[
\frac{\delta f_i} {\Delta f}
\frac{2\mathcal R_{12}^2(f_i)  }{S_{1n}(f_i)S_{2n}(f_i)}
\r]
\delta_{ij}.
\el
We  plot $ F_{ii}(f) $ of (\ref{FisherCorNew}) and
of   (\ref{FisherCorNewNoise}) in Fig.\ref{FisherDiaSeto}.
They differ significantly at high frequencies.
Thus, we shall use the full expression (\ref{FisherCorNew})
in  computations later.

Given $ F_{ij}$,
one solves Eq.(\ref {trialCorNew}) for the estimated spectrum
\be\label{SpectrEstimCorNew1}
S_h(f_i)
= S^{(0)}_h(f_i)
     -\sum_{j=1}^{N} ( F^{\,-1})_{ij}
     \frac{\partial\mathcal L} {\partial S_{h}(f_j)}\bigg|_{S^{(0)}_h } .
\ee
To avoid random outcomes from one set of datastream,
similar to (\ref{samplemean}),
we shall replace $ \frac{\partial\mathcal L} {\partial S_{h}(f_k)}$ by its  sample mean
in practical computation
\be\label{SpectrEstimCorNew2}
S_h(f_i)
= S^{(0)}_h(f_i)
     - \l(
\frac{\delta f_i} {\Delta f}
\frac{2\mathcal R_{12}^2(f_i)  }{M(f_i)}
+\left.\frac{2 N^2( f_i)}{S^2_h(f_i)M^2(f_i)}
\r)^{-1}
  \l\langle
     \frac{\partial\mathcal L} {\partial S_{h}(f_i)}
     \r\rangle_t
      \right|_{S^{(0)}_h } \, ,
\ee
where the expression (\ref{FisherCorNew}) has been  used.
This equation will be used to estimate the spectrum of RGW numerically by
Newton-Raphson iteration \cite{Oh1999,Hinshaw2003Apjs,Press1992}.

\begin{figure}[htbp]
\centering
  \includegraphics[width=0.8\columnwidth]{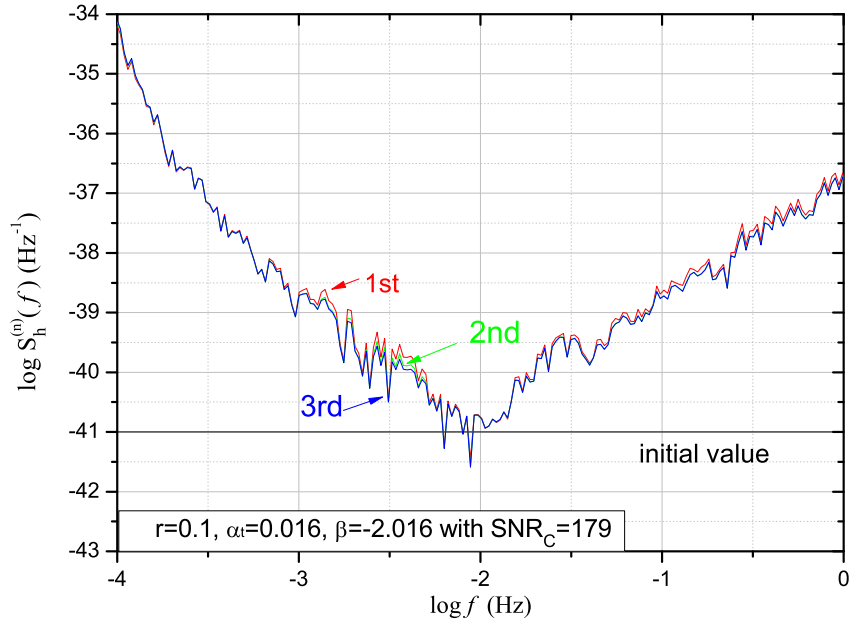}\\
  \caption{
  The  estimator of spectrum $S_h^{(n)}(f)$ in each iterative step for a pair.
  }\label{SpecEstIterat}
\end{figure}

\begin{figure}[htbp]
\centering
  \includegraphics[width=0.8\columnwidth]{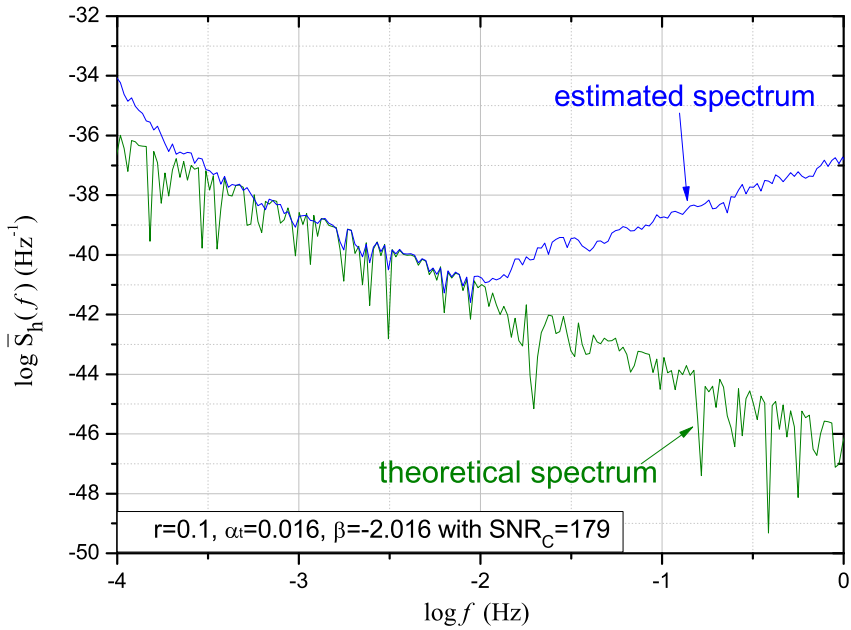}\\
  \caption{
  The estimated spectrum by the correlation method for a pair.
  The  theoretical spectrum is also shown.
  }\label{ShEstCorSeto}
\end{figure}

We perform a  simulation to estimate $S_h(f)$.
We divide a one-year duration     into 100 periods,
which are regarded as $n=100$   realizations of data output.
Thus,
one  observation period $T_1\simeq1$year$ / 100 \simeq 3.2\times10^{5}$s.
The working frequency range $(10^{-4} \sim 1)$ Hz
is divided into  $N \simeq  10637$ segments
and the width of each segment is $\delta f\simeq 9.4\times 10^{-5}$\,Hz.
Thus, each segment contains $\delta f/\Delta f\simeq 30$ frequency points.
For  RGW   we take   $r=0.1$, $\alpha_t=0.016$ and $\beta=-2.016$,
and the formula (\ref{SNRC}) yields  SNR$_{C}=179$.
We numerically generate the output response
$[\tilde h_{1, 1}(f_i),\cdots, \tilde h_{1, \, n}(f_i); \tilde h_{2, 1}(f_i),\cdots,
\tilde h_{2, \, n}(f_i)]$ according to (\ref{PDFRGW})
and the noise $[\tilde n_{1, 1}(f_i),\cdots, \tilde n_{1, \, n}(f_i);
   \tilde n_{2, 1}(f_i),\cdots, \tilde n_{2, \, n}(f_i)]$  according to (\ref{PDFNoise})
of a pair, for $i=1,\cdots,10637$ and for $n=100$ realizations.
We use Eq.~(\ref{corZi}) to calculate the correlated signal
$[Z_{i, 1},\cdots,Z_{i,\, n}]$ for $n=100$ realizations.
Using these generated   data streams,
we apply Newton-Raphson iteration  to Eq.~(\ref{SpectrEstimCorNew2})
to estimate the spectrum of RGW numerically.
Fig.~\ref{SpecEstIterat} shows the resulting  estimator of spectrum
$S_h^{(n)}(f)$ after each iterative step.
It is seen that after three iterations,
$S_h^{(n)}(f)$ converges.
The estimated RGW spectrum $\bar S_h(f_i)$ is shown in
in Fig.~\ref{ShEstCorSeto}.
For illustration,
the theoretical spectrum is also shown.
It is seen that the rapid oscillations in the estimated spectrum
are  smoother than the theoretical one.
This is because
the estimated spectrum is actually
a mean $\sim\frac{1}{\delta f_i} \sum_{f\in F_i} \Delta f$
in each segment in this method.

We compare the correlation method in this section
with the ensemble averaging method of Section 7.
Firstly,  the  correlation method uses
the average value as the representative for each segment,
and losses some fine information about the RGW spectrum.
Secondly, this method needs more computing time
 by the iteration for each frequency point.
However, this method can estimate the parameters directly
as in the following subsection
whereas  the ensemble averaging method can not.

\subsection{Parameter estimation}

Now  we estimate parameters of RGW
by using the correlated data stream $Z_{\, i} $
in a Bayesian approach.
Consider   the PDF  as in Eq.~(\ref{PDFcorNew}),
\be\label{BayesPDFnew}
f ({\bf Z};\bm\theta) =
\frac{1}{(2\pi  )^{\frac{N}{2}}\, ( \Pi_i^N \sigma_i^2(\bm\theta))^{\frac12 } }
\exp\l\{ -\frac{1}{2}
\sum_i^N { \frac{ \l( Z_i - \mu_i (\bm\theta) \r)^2 }{\sigma_i^2(\bm\theta)} }
    \r\} \, ,
\ee
where $\mu_i (\bm\theta)$ and $\sigma_i^2(\bm\theta)$
given by (\ref{MeanCorNew}) and (\ref{VarCorNew})
respectively,
are now regarded as functions of  parameters $\bm \theta$
through  the theoretical spectrum $S_h(f)$.
The likelihood function $ \mathcal L= - \ln f( \bm\theta) $ can also be
Taylor expanded  around certain values $  \bm {\bar \theta}$
\[
\mathcal L = \bar{ \mathcal L}
+\sum_a  \frac{\partial\mathcal L}{\partial\theta_a}  \bigg|_{\bm {\bar \theta}} \,
           (\theta_a -\bar \theta_a)
+\frac12 \sum_{a,b} \,
\frac{\partial^2\mathcal L}{\partial \theta_a  \partial \theta_b }  \bigg|_{\bm {\bar \theta}}
(\theta_a -\bar \theta_a)(\theta_b -\bar \theta_b)+\,   \cdots \, .
\]
Now we require   $  \bm {\bar \theta}$ to be
the ML estimator,
at which
\be \label{pL}
\frac{\partial \mathcal L }{\partial\theta_a}
\,  \bigg|_{ \bm {\bar \theta} }
    =0,  ~~~~a=1,2,3.
\ee
Based on (\ref{pL}),
we use Newton-Raphson method  \cite{Oh1999,Hinshaw2003Apjs,Press1992}
to estimate $\bar \theta$.
The first order derivative is expanded
around the trial $\bm \theta ^{(0)}$
 as the following
\be \label{trialthetaCorNew}
\frac{\partial\mathcal L}{\partial \theta_a}\bigg|_{\bm \theta  }
\simeq
\frac{\partial\mathcal L}{\partial \theta_a}\bigg|_{\bm \theta ^{(0)}  }
+     \sum_{b=1}^{3}\frac{\partial ^2 \mathcal L}{ \partial  \theta_a \partial \theta_b }
  \bigg|_{ \bm \theta ^{(0)} }
   \l(\theta_{b} -\theta_b^{(0)}  \r) =0,
       ~~~~ a=1,2,3.
\ee
The second order derivative in the above is approximately replaced
by  the Fisher matrix,  leading to
\be \label{fstord0NewCor}
\frac{\partial\mathcal L}{\partial \theta_a}\bigg|_{\bm \theta ^{(0)}  }
+     \sum_{b=1}^{3}F_{ab}
  \bigg|_{ \bm \theta ^{(0)} }
   \l(\theta_{b} -\theta_b^{(0)}  \r) =0,
       ~~~~ a=1,2,3 ,
\ee
from which  we obtain
\be\label{EstThetaCorNew2}
\theta_{a} = \theta_a^{(0)}
- \sum_{b=1}^{3}
F_{ab}^{-1}
\l\langle
\frac{\partial\mathcal L}{\partial \theta_b}
\r\rangle_t
   \,    \bigg|_{\bm \theta ^{(0)}  },  ~~~~ a=1,2,3 .
\ee
By iteration,
one can obtain an estimate of the parameters $\{\theta_{a}\}$.
The explicit expressions of
 derivatives   in the above
are given by the chain rule by using (\ref{LikePShCorNew}),
\bl\label{LikliThetaCorNew}
\frac{\partial\mathcal L}{\partial \theta_a}
=&
\sum_{i=1}^N   \frac{\partial  \mathcal L}{\partial  S_{h}(f_i) }
    \frac{\partial  S_h(f_i)}{\partial \theta_a}
\nn\\
= &
\sum_{i=1}^N
\Bigg[
 \frac{N(f_i)}{S_{h}(f_i)M(f_i)}
-\frac{ 8\l( Z_i - (\frac{\delta f_i} {2\Delta f} S_h(f_i)\mathcal R_{12}(f_i)  )  \r)^2 }
{\frac{\delta f_i} {\Delta f} S_{h}(f_i)\l[M(f_i)\r]^2} N(f_i)
\nn\\
&
-\frac{4 \mathcal R_{12}(f_i)\l( Z_i - (\frac{\delta f_i} {2\Delta f} S_h(f_i)\mathcal R_{12}(f_i)  ) \r) }{M(f_i)}
\Bigg]
\frac{\partial  S_h(f_i)}{\partial \theta_a}
,  ~~~~ a=1,2,3
\el
The Fisher matrix is provided by the following
\bl
F_{ab} =
 \l\langle\frac{\partial ^2\mathcal L} {\partial  \theta_a\partial \theta_b}\r\rangle
=&
\sum_{k,l=1}^N\frac{\partial  \mu_k}{\partial   \theta_a}
\frac{\delta_{kl}}{\sigma_k^2}\frac{\partial   \mu_l}{\partial  \theta_b}
+\frac12\sum_{k,l,m,r=1}^N\l(\frac{\delta_{kl}}{\sigma_k^2}
\delta_{lm}\frac{\partial   (\sigma_l^2)}{\partial \theta_a}
\frac{\delta_{mr}}{\sigma_m^2}
\delta_{rk}\frac{\partial  (\sigma_r^2)}{\partial  \theta_b}\r)
,  ~~~~ a=1,2,3
\el
Using  (\ref{MeanCorNew}) (\ref{VarCorNew}) and the chain rule,
 one has
\bl\label{FisherParaNew}
F_{ab} =
\sum_{k}^N
\frac{\delta f_k} {\Delta f}\frac{2\mathcal R_{12}^2(f_k)}{M(f_k)}
\frac{\partial S_h(f_k)}{\partial   \theta_a}
\frac{\partial S_h(f_k)}{\partial  \theta_b}
+\sum_{k=1}^N
\l(\frac{2N^2(f_k)}{S_h^2(f_k)M^2(f_k)}
\frac{\partial S_h(f_k)}{\partial \theta_a}
\frac{\partial S_h(f_k)}{\partial  \theta_b}\r).
\el
which is not degenerate,
  since $\mu_i$ and $\sigma_i^2$, in
(\ref{MeanCorNew}) and (\ref{VarNoiseCorNew}) respectively,
carry  information on frequency $f_i$.
Thus,  (\ref{EstThetaCorNew2}) can be used to estimate
the three parameters at the same time.
In the limit of dominating noise,
one has $M(f)\simeq S_{n1}(f)S_{n2}(f)$
and $N(f)\simeq0$,
and (\ref{FisherParaNew}) reduces to
\bl
F_{ab} =
\sum_{k}^N
\frac{\delta f_k} {\Delta f}\frac{2\mathcal R_{12}^2(f_k) }{ S_{n1}(f_k)S_{n2}(f_k)}
\frac{\partial S_h(f_k)}{\partial   \theta_a}
\frac{\partial S_h(f_k)}{\partial  \theta_b}.
\el
Replacing $\Delta f$ with $1/T_1$,
and   $\sum_{i=1}^N\delta f$ with   $\int_0^\infty df$,
the  above becomes
\bl
F_{ab} =
T_1  \int_0^\infty df\frac{2\mathcal R_{12}^2(f)  }{ S_{n1}(f)S_{n2}(f)}
\frac{\partial S_h(f)}{\partial   \theta_a}
\frac{\partial S_h(f)}{\partial  \theta_b},
\el
which agrees with  (3.11) in Ref.\cite{Seto2006}.

\begin{figure}[htbp]
\centering
  \includegraphics[width=0.6\columnwidth]{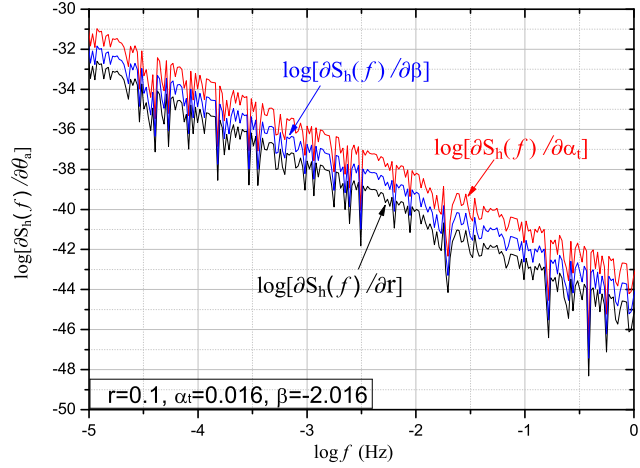}\\
  \caption{
  $\partial S_h /\partial\theta_a$
   for  $r=0.1,\beta=-2.016$ and $\alpha_t=0.016$
   }\label{PartialShp}
\end{figure}

The element $F_{ab}$  can be viewed
as an ``inner"  product of two vectors
$\frac{\partial S_h }{\partial\theta_a}$ and
 $\frac{\partial S_h }{\partial\theta_b}$.
When the  vectors $\frac{\partial S_h(f)}{\partial\theta_a}$ with $a=1,2,3$
are orthogonal
 to each other, $F_{ab}$ will be diagonal,
 and the errors in estimates of different parameters will be uncorrelated.
On the other hand,
when two $\frac{\partial S_h }{\partial\theta_a}$ and
$\frac{\partial S_h }{\partial\theta_b}$ have similar shapes,
 $\theta_a$ and $\theta_b$ will be  degenerate,
and their effects are  difficult to distinguish in estimation.
In Fig.\ref{PartialShp},
the three curves $\frac{\partial S_h(f)}{\partial\theta_a}$
 based on the theoretical spectrum
  are plotted,
showing  that the three parameters of RGW  have strong  degeneracy
within a small frequency range.

For    $r=0.1$, $\alpha_t=0.016$ and $\beta=-2.016$ with SNR$_{C}$=179,
we  use the data streams generated in Section 8.2
and use (\ref{EstThetaCorNew2}) to numerically estimate $r$,
 and find  that $r$ converges to $r_{ML}=0.1070$
 after nine iterations.

As discussed before,
in the neighborhood   of $\bm {\bar  \theta} $, one has
the following Bayesian PDF
in the parameter space
\be\label{ThetaBayesian}
f ( \bm \theta)  \propto    \exp{[-\mathcal L] }
 \propto
\exp\l[-\frac12( \bm\theta-\bm{ \bar \theta} )
 \bm F (\bm\theta -\bm{ \bar \theta} )^T\r] \, .
\ee
The resolution of the parameters will  be
\be
\Delta\theta_a
= 1.96\sigma_{\theta_a}
= 1.96 \sqrt {\l(\bm F^{-1}\r)_{aa}}
\ \  {\rm\ at\ } 95\% \,\,{\rm cl} \, ,
\ee
and the correlation coefficient between two
parameters will be
\be
CR_{\theta_a\theta_b}\equiv
\frac{\Big\langle \Big(\theta_a-\langle\theta_a\rangle\Big)
    \Big(\theta_b-\langle\theta_b\rangle\Big)\Big\rangle}
    {\sigma_{\theta_a}\sigma_{\theta_a}}
    = \frac{\l(\bm F^{-1}\r)_{ab}}
    { \sqrt {\l(\bm F^{-1}\r)_{aa}\l(\bm F^{-1}\r)_{bb}}}\,.
\ee
Note that $CR_{\theta_a\theta_b}=0$ indicates the independency of
$\theta_a$ and $\theta_b$,
and $|CR_{\theta_a\theta_b}|=1$ indicates the complete correlation of
$\theta_a$ and $\theta_b$.
Comparing Table \ref{resolution abrSeto} and Table \ref{resolution abr},
when $r$, $\alpha_t$ and $\beta$ are estimated at the same time,
where Table \ref{resolution abrSeto} lists
 the resolutions, correlations and the corresponding values of SNR$_{12}$ ($\geq 3.29$),
we  find that when estimating the three parameters simultaneously,
the resolution would get worse.
This is due to the degeneracy of the three parameters
as shown in Fig.~\ref{PartialShp},
which can also be seen with $|CR_{\theta_a\theta_b}|\simeq1$
in Table \ref{resolution abrSeto}.
Since the amplitude of the spectrum increases with $r$, $\beta$ and $\alpha_t$,
when simultaneously
estimating the three parameters,
a larger estimated value of $\beta$ than its true value
will lead to smaller estimated $r$ and $\alpha_t$ than their true values,
or vice versa.
This is reflected in the negative signs of
$CR_{r\beta}$ and $CR_{\alpha_t\beta}$ in Table \ref{resolution abrSeto}.
 Besides,
we find that if estimating only two parameters with the third one fixed,
the correlation coefficients between every two parameters
are all negative, which is also an  expected feature.
\begin{table}[htbp]
\caption{Resolution of $r$,
$\alpha_t$ and $\beta$ at $95\%$ cl for a pair,
and correlations between them}
\begin{center}
\label{resolution abrSeto}
\begin{tabular}{|c |c|c|| c| c|c||c|c|c||c|}
\hline
$r$&$\alpha_t$ & $\beta$ & $\Delta r/r$&$\Delta\alpha_t$ & $\Delta\beta$
& $CR_{r\alpha_t}$& $CR_{r\beta}$ & $CR_{\alpha_t\beta}$
&SNR$_{C}$
\\
\hline
0.05&0.01&-2.016&
$1.54\times10^{4}$  & $26.3$ &$450$  & $0.999697$& $-0.999923$ & $-0.999925$& 4.36
\\
\hline
0.05&0.015&-2.016&
$997$  & $1.70$ &$29.1$  & $0.999724$& $-0.999930$ & $-0.999932$& 72.8
\\
\hline
0.1&0   & -1.93 &
$6.40\times10^{3}$& $11.0$ &$188$  & $0.999634$& $-0.999907$ & $-0.999910$&  8.34
\\
\hline
0.1&0.01&-2.016&
$7.71\times10^{3}$  & $13.2$ &$226$  & $0.999696$& $-0.999923$ & $-0.999925$& 8.62
\\
\hline
0.1&0.01& -1.93&
$67.5$  & $0.117$ &$1.99$  & $0.999460$& $-0.999863$ & $-0.999867$&  390
\\
\hline
0.1&0.016&-2.016&
$329$  & $0.561$ &$9.60$ & $0.999681$& $-0.999919$ & $-0.999921$ & 179
\\
\hline
\end{tabular}
\end{center}
\end{table}

\section{Conclusion }

We have presented  a study of statistical signal processing
for the  RGW detection by space-borne interferometers,
using  LISA as an example,
and have shown how to estimate the RGW spectrum and parameters
from the output signals in the future.
We have given the relevant  formulations of estimations,
which apply to LISA, as well as to other space-borne interferometers
with some appropriate modifications.

For a single  interferometer,
the Michelson  is shown to have a better sensitivity
than Sagnac, and symmetrized Sagnac,
due to its greater transfer function $\mathcal{R}$,
even though its noise is larger.
A pair  has the advantage of suppressing
the noise level by cross-correlation,
so that RGW signal in the cross-correlated output
will be accumulating with observation time,
leading to  a higher sensitivity than a single case.
We have given the expressions of SNR   for both a single and a pair,
which are  $4\sim 6$ orders of magnitude
higher  than those  of ground-based ones for the default RGW parameters.
We have shown that a single is not practical to estimate the RGW spectrum
 when  noise is dominantly large,
because we do not know the precise  noise that actually occurs
in the data.

For a pair of interferometers,
we have used the  cross-correlated integrated signals $C$ in (\ref{correlation signal})
and calculated SNR$_{12}$ as in (\ref{SNR max correlation})
which provides a statistical criterion for detection of RGW.
However,
one cannot estimate the spectrum by the integrated signals $C$,
  because it is an integration over frequency.
Assuming Gaussian output signals,
and we have calculated the covariance  of signals,
obtained  the Gaussian PDF, the likelihood function $\mathcal L$,
and the Fisher matrix.
By the Bayesian approach,
we estimate one parameter  and compute the resolution  using $C$.
In the second method,
we have proposed applying the ensemble averaging
method to estimate the spectrum,
  using the    un-integrated output signals of a pair.
(\ref{s1s2toSpectrum4}) is the main formula.
We have demonstrated by simulation that the method will be effective
in estimating the spectrum.
Besides,
the method is simple and does not depend on detailed knowledge of  the noise.
For the third method,
we have also studied the correlation variable of un-integrated signals
from (\ref{corZi}).
We have obtained the formulations
for estimation of the spectrum and parameters of RGW by the ML-estimation.
Eqs.~(\ref{SpectrEstimCorNew2}) and (\ref{EstThetaCorNew2})
are the main formulae.
We have shown  by   simulations
 that the method is feasible
when the data set is sufficiently large.
This method uses the mean value for each segment
and thus loses some fine information on the RGW spectrum,
but it is capable of estimating all three parameters.

There are other effects to be analyzed
that are not presented in this paper.
In particular,
other types of GWs that are different from  RGW also exist in the Universe,
and should be separated in order to estimate the RGW spectrum.
GWs from a resolved astrophysical source, either periodic or short-lived,
 can be distinguished in principle.
The real concern is the stochastic foreground that may
by mixed with RGW.
So far, the theoretical spectrum of this foreground
is less known and highly model-dependent.
For a definite model of the foreground spectrum,
Ref.~\cite{AdamsCornish2014} discusses a discrimination method
using the spectral shapes and the time modulation of the signal.
The  estimation of RGW  in the presence of  foreground
will require substantial analysis
and will be left for future work.

\

\textbf{Acknowledgements}

Y. Zhang is supported by
NSFC Grant No. 11421303, 11675165, 11633001
SRFDP and CAS the Strategic Priority Research Program
``The Emergence of Cosmological Structures"
of the Chinese Academy of Sciences, Grant No. XDB09000000.

\numberwithin{equation}{section}

\appendix

\section{ The Fisher matrix       for a pair}

Given the data   of cross-correlated signals  $\bm C$  in (\ref{Ci})
for a pair,
 we assume that the PDF is multivariate Gaussian
\be\label{pdfmultapp}
f({\bf C}  )=
\frac{1}{(2\pi)^{\frac{N}{2}}\text{det}^{\frac12}[\bm \Sigma ] }
\exp\l\{-\frac12 { (\bm C-\bm\mu) }\ \bm\Sigma^{-1}
 \   \l(\bm C-\bm\mu   \r) ^T\r\} \, ,
\ee
where the mean  $\mu_i$ and covariance matrix $\Sigma_{ij}$
are given by  (\ref{mui2}) and (\ref{sigmaij2}) respectively,
both being functions of the spectrum $S_h(f)$.
The likelihood function  is
(dropping an irrelevant   constant $\frac12 N \ln 2\pi$)
\be\label{likeliapp}
 \mathcal L\equiv -  \ln f= \frac12 \ln \text{det} [\bm \Sigma ]
 +\frac12 { (\bm C-\bm\mu) }\ \bm\Sigma^{-1}
 \   \l(\bm C-\bm\mu   \r) ^T .
\ee
The first order derivative is \cite{Kay}
\bl\label{PartialLikelyhood1}
\frac{\delta  \mathcal L}
{\delta   S_{h}(f)}
=&\frac12\text{tr}\l(\bm\Sigma^{-1}
\frac{\delta  \bm\Sigma}{\delta  S_{h}(f)}\r)
-\frac12{\bm y}\
\bm\Sigma^{-1}
\frac{\delta  \bm\Sigma}{\delta  S_{h}(f)}
\bm\Sigma^{-1}\ {\bm y}^T
-\frac{ \delta \bm\mu}{\delta  S_{h}(f)}
\bm\Sigma^{-1}\ {\bm y}^T
\el
where   $\bm y    \equiv\bm C-\bm\mu $,
and the second order derivative is
\bl\label{PartialLikelyhood2}
\frac{\delta ^2\mathcal L}
{\delta  S_{h}(f)\delta  S_h(   f' )  }
=&-\frac12\text{tr}\l(\bm\Sigma^{-1}
\frac{\delta \bm\Sigma}{\delta  S_{h}( f')}
\bm\Sigma^{-1}
\frac{\delta \bm\Sigma}{\delta  S_{h}(f)}\r)
+\frac12\text{tr}\l(\bm\Sigma^{-1}
\frac{\delta ^2\bm\Sigma}{\delta  S_{h}( f')\delta S_{h}(f)}\r)
\nn\\
&
+{\bm y}\bm\Sigma^{-1}
\frac{\delta  \bm\Sigma}{\delta  S_{h}( f')}
\bm\Sigma^{-1}
\frac{\delta \bm\Sigma}{\delta  S_{h}(f)}
\bm\Sigma^{-1}\ {\bm y}^T
+\frac{\delta \bm\mu}{\delta  S_h( f')} \
\bm\Sigma^{-1}
\frac{\delta \bm\Sigma}{\delta  S_{h}(f)}
\bm\Sigma^{-1}\ {\bm y}^T
\nn\\
&
-\frac12{\bm y}\
\bm\Sigma^{-1}
\frac{\delta ^2\bm\Sigma}{\delta  S_{h}( f')\delta  S_{h}(f)}
\bm\Sigma^{-1}\ {\bm y}^T
-\frac{\delta ^2\bm\mu}{ \delta  S_{h}( f') \delta  S_{h}(f)}
\bm\Sigma^{-1}\ {\bm y}^T
\nn\\
&
+\frac{\delta  \bm\mu}{\delta  S_{h}(f)}
\bm\Sigma^{-1}\frac{\delta \bm\mu^T}{\delta  S_{h}( f')}
+\frac{\delta  \bm\mu}{\delta  S_{h}(f)}
\bm\Sigma^{-1}
\frac{\delta \bm\Sigma}{\delta  S_{h}( f')}
\bm\Sigma^{-1}\ {\bm y}^T \, .
\el
Taking the expectated value of the above yields   the Fisher matrix
\bl\label{Likelyhood2}
 \mathcal F( f, f'   ) =
 \l\langle\frac{\delta ^2\mathcal L} {\delta  S_{h}(f)\delta   S_h( f')}\r\rangle
=&\frac{\delta  \bm\mu}{\delta   S_{h}(f)}
\bm\Sigma^{-1}\frac{\delta  \bm\mu^T}{\delta  S_{h}( f')}
+\frac12\text{tr}\l(\bm\Sigma^{-1}
\frac{\delta  \bm\Sigma}{\delta  S_{h}( f')}
\bm\Sigma^{-1}
\frac{\delta \bm\Sigma}{\delta  S_{h}(f)}\r).
\el
where $\l\langle{\bm y}^T\r\rangle=0$
and $\l\langle{\bm y}^T{\bm y}\r\rangle=\bm\Sigma$ are used.
Using  $\Sigma_{ij}$  of (\ref{sigmaij2}),
 one has
\bl \label{pSh}
\frac{\delta  \mathcal L} {\delta  S_{h}(f)}
= & \sum^N_{l}
 \bigg[  \frac{1}{2 \mu_l} -\frac{C^2_i}{2 b \mu_i^2} +\frac{1}{2 b}  \bigg]
   \frac{\delta  \mu_l}{\delta  S_{h}(f)},
\el
\bl \label{Fcon}
 \mathcal F( f, f'   )
=&  \sum_l^N  \left( \frac{1}{ b\mu_l }  +  \frac{1}{2 \mu_l^2}\right)
  \frac{\delta  \mu_l}{\delta  S_{h}( f')}
   \frac{\delta  \mu_l}{\delta  S_{h}(f)} \, ,
\el
where  $b \equiv\frac{5 }{3 } $.
By (\ref{mui2}),
the derivative of $\mu$  is given by
\bl \label{mufunctional}
\frac{\delta  \mu_i}{\delta  S_{h}(f)}
 = \frac{T_i}{2 b} \frac{\delta  m}{\delta  S_{h}(f)}
 =   \frac{T_i}{2 b}
  \frac{ S_h(f)\, \gamma^2(f)}{M(f)}
  \l( 1-  \frac{N(f) }{M(f)}  \r)
  ,~~~~~~~ i=1,\cdots,N,
\el
where $M(f)$ is defined by  (\ref{MS12}) and
\be \label{N}
N(f)\equiv \frac12 S_h(f) \frac{\partial{M(f)}} {\partial S_h(f)}
   =  \frac12  \Big[ S_{1n}(f)+S_{2n}(f) \Big]  \mathcal R(f) S_h(f)
    + \l[ \mathcal R^2(f) +\mathcal R_{12}^2(f) \r] S^2 _h (f) .
\ee
Using   (\ref {mui2}) and (\ref{mufunctional}),
 the first order  derivative is
\bl\label{delL}
\frac{\delta  \mathcal L}{\delta  S_{h}(f)}
= & \frac12 \frac{ S_h(f)\, \gamma^2(f)}{M(f)}   \l( 1- \frac{N(f) }{M(f)} \r)
 \l( \frac{N}{m}   - \frac{2}{m^2} \sum^N_{i} \frac{C^2_i }{T_i}
                +  \frac{1}{2 b^2} \sum^N_{i}  T_i \r) ,
\el
and  the Fisher matrix is
\bl \label{Fcontinuous}
\mathcal F( f, f'   )
=&
 \l[ \frac{ S_h( f')\, \gamma^2( f')}{M( f^{\,'})}
  \l( 1-  \frac{N( f') }{M( f')}  \r)\r]
  \l[\frac{ S_h(f)\, \gamma^2(f)}{M(f)}
  \l( 1-  \frac{N(f) }{M(f)}  \r)\r] \frac{1}{2}
  \l(  \frac{N}{m^2} +  \frac{ 9}{25 m } \sum_l^N T_l \r)\,.
\el
From the above formulae,
we also derive the Fisher matrix $F_{ab}$ for parameter estimation.
Consider    the PDF in (\ref {pdfmultapp})
\be \label{PDFthe}
f({\bf C};\bm\theta)=
\frac{1}{(2\pi)^{\frac{N}{2}}\text{det}^{\frac12}[\bm \Sigma(\bm\theta)] }
\exp\l\{-\frac12 {\l(\bm C-\bm\mu(\bm\theta)\r)}\ \bm\Sigma^{-1}
(\bm\theta)\ {\l(\bm C-\bm\mu(\bm\theta)\r)}^T\r\} \, ,
\ee
where  $\mu(\theta)$ and  $\bm \Sigma(\bm\theta)$
now depend  on the RGW parameters  $\bm \theta = (r,  \beta, \alpha_t)$
through the theoretical spectrum $S_h(f)$
 in (\ref{mui2}) and (\ref{sigmaij2}).
Using  the result  (\ref {pSh}) (\ref{Fcon}), by the chain rule,
one obtains  the derivatives   with respect to the parameters
\be \label{1st0param}
\frac{\partial \mathcal L}{\partial \theta_a}
=  \frac12\sum^N_{i}
 \bigg[  \frac{1}{ \mu_i} -\frac{C^2_i}{ b \mu_i^2} +\frac{1}{b}  \bigg]
 \frac{\partial    \mu_i}{\partial \theta_a} \, ,
\ee
\be\label{Likelyhoodtheta}
\mathcal F_{ab} = \l\langle \frac{\partial^2\mathcal L}
  {\partial \theta_a  \partial \theta_b }\r\rangle
=  \sum_i^N  \l( \frac{1}{ b\mu_i }  +  \frac{1}{2 \mu_i^2}\r)
   \frac{\partial   \mu_i} {\partial  \theta_a}
   \frac{\partial   \mu_i} {\partial  \theta_b} \, .
\ee
Taking derivative of $\mu$ in Eq.~(\ref{mui2}) with respect to $\theta_a$
yields
\bl\label{partialmu}
    \frac{\partial\mu_i}{\partial\theta_a}
= & \frac{T_i}{A}
  \int_{0}^\infty df
                \frac{ S_h(f)\, \gamma^2(f)}{M(f)}
                \l( 1- \frac{N(f) }{M(f)} \r)\,
                   \frac{\partial  S_{h}(f)}{\partial  \theta_a},
       ~~~~ a=1,2,3, ~~~ i=1,\cdots,N.
\el
Substituting (\ref{partialmu})
into (\ref{1st0param}) and (\ref{Likelyhoodtheta})
leads to   (\ref{parameq})  and   (\ref{FFish}).

\end{document}